\documentclass[11pt, final]{article}
\usepackage{MRnotes}

\title{Effective dynamics of open 2D CFTs}
\author{Lorenzo Toni}

\affiliation{
	Center for Quantum Mathematics and Physics (QMAP)\\
	Department of Physics \& Astronomy, University of California, Davis, CA 95616 USA}

\emailAdd{ltoni@ucdavis.edu}

\abstract{We analyze the momentum-space representations of causal response functions of scalar primary operators in an arbitrary 2D CFT, which can be exploited to characterize the effective dynamics of open quantum CFTs. We derive our results by analytically continuing Euclidean correlators to Lorentzian signature. 

While the two- and three-point functions can be continued straightforwardly, the case of the four-point function is more challenging. In principle, the analytic continuation of the four-point function can be performed in full generality using either the Virasoro conformal block expansion or by mapping the conformal blocks to radial coordinates. However, these frameworks are mathematically intractable for explicit momentum-space calculations. We argue that the global block expansion offers a valuable alternative, as it allows us to recover the time-ordered correlators while preserving a relatively simple analytic structure. From these causal configurations, the four-point response function is systematically constructed and its momentum-space representation is evaluated.
}

\begin{document} 
\maketitle


\section{Introduction}\label{sec:intro}

An open quantum system is one that interacts with an external quantum (or classical) system. The latter is usually referred to as environment or bath. In quantum mechanics and QFT, many problems arise in which several systems are coupled together, but one or more of these systems are not of primary interest. These systems can be regarded as the environment with which the others, commonly referred to as probe systems, interact. Since we don't usually have any control over the external bath, we would like to obtain an effective description of the full system in terms of the degrees of freedom of the probe system only, the effect of the environment being encoded in the higher order corrections to the dynamics of the probe system~\cite{Feynman:1963}. Within this framework, we can describe a broad spectrum of phenomena, ranging from quantum dissipation, non-unitary evolution and decoherence \cite{Caldeira:1983, Zurek:2003, Grozdanov:2015, Boyanovsky:2015, Baidya:2017, Boyanovsky:2018, Agon:2018}. 

The study of open quantum systems inherently necessitates the use of mixed states to track the generation of entanglement between the probe system and its surrounding environment. This out-of-equilibrium dynamics is naturally formulated within the Schwinger-Keldysh formalism, which systematically doubles the physical degrees of freedom to account for the environment's influence \cite{Schwinger:1961, Keldysh:1965}. This doubling reflects the structure of time evolution along a closed time path contour in the complex time plane, where the system evolves forward in time and then backward, ensuring a consistent treatment of expectation values within the density matrix \cite{Landsman:1987, Chou:1985, Breuer:2002, Kamenev:2009, Kamenev_2011, Sieberer:2016}. Consequently, causality and unitarity are manifest at the level of the formalism. See \cite{Haehl:2017, Haehl:2017_2} for a modern introduction.

However, we still lack a full understanding of the general dynamics of open quantum systems. If the coupling between the probe system and the environment is weak, we can exploit the well-known machinery of perturbation theory. In general, the resulting effective field theory (EFT) is non-local, the reason being that the full derivation of the EFT is intrinsically a non-perturbative question. Indeed, from a microscopic point of view, a local description is only valid at time scales larger than the environmental memory time scale, which is the time the environment fields take to forget the information about the configuration they started from. Since this time scale is inversely proportional to the strength of the interaction between the probe system and the environment, one has to wait a sufficiently long time for a local description to emerge. Inevitably, we have to deal with non-local interaction terms when we employ standard perturbation theory in the study of open quantum systems. 

In this paper we consider both the probe system and the bath to be 2D CFTs, leaving the coupling term between them as arbitrary as possible. Because of the conformal symmetry shared by both systems, we can make sense of the infinite set of terms that arise in the corresponding EFT. In particular, we can explicitly compute the lower-order corrections to the dynamics of the probe system in momentum space in the form of causal response functions involving the bath degrees of freedom only.

Although we have so far implicitly assumed that our open system is at zero temperature, it is more realistic to model the environment as a thermal bath in contact with the probe system. For our purposes, we consider both CFTs describing the probe system and the environment to be at finite temperature. Consequently, the lower-order corrections within the resulting EFT will involve the Fourier transforms of the thermal two-point, three-point and four-point response functions. While the momentum-space structure of CFT correlators is well-documented \cite{Gillioz:2020, Gillioz:20202, Gillioz:2021, Gillioz:2025, Bautista:2020} and explicit expressions for the thermal two-point~\cite{Son:2002} and three-point functions~\cite{Becker:2014} are well established, the case of the thermal four-point function is still an open problem. The main result of this paper is the computation of the Fourier transform of the thermal four-point function of identical scalar fields. 

A remarkable fact about thermal states is that the real-time response functions are related to fluctuations about equilibrium via fluctuation-dissipation theorems. These theorems originate from the KMS relations \cite{Martin:1959, Kubo:1957, Chou:1985}, which impose a structural periodicity on thermal correlators under imaginary translations of the time coordinates. Fluctuation-dissipation relations are traditionally expressed by relating the retarded Green’s function in thermal equilibrium (causal response) to the symmetrized two-point function (the fluctuation), but they clearly admit generalization to higher-point functions in terms of fully nested commutators and anti-commutators \cite{Haehl:2017_kms, Gransee:2017}. Crucially, the complete set of KMS constraints simplifies considerably when the formalism is broadened to encompass both time-ordered and out-of-time-order (OTO) \cite{Haehl:2019} correlation functions. Using these generalizations, a causal basis was constructed in \cite{Haehl:2017_kms} for the full set of $n$-point thermal correlators. This basis is spanned by nested commutators where the position of the innermost operator is held fixed. Within this framework, any physical response function can be reconstructed as a linear combination of these basis elements modulated by appropriate step functions.

However, constructing these response functions introduces a subtle technical issue. Our open quantum system is defined on a 2D Lorentzian spacetime, whereas CFTs are usually discussed in Euclidean signature. In principle, Lorentzian correlators can be recovered from their Euclidean counterparts via analytic continuation \cite{Haag:1967, Hartman:2019}. While this analytic continuation is straightforward for the two- and three-point functions, the four-point function requires a more careful analysis.

Indeed, the four-point function is usually decomposed as a sum over conformal blocks. These blocks can be expressed in several representations, each possessing different analytic properties and domains of convergence. In particular, the analytic continuation of the four-point function can be performed in full generality using either the Virasoro conformal block expansion \cite{Zamolodchikov:1987} or by mapping the conformal blocks to radial coordinates \cite{Hogervorst:2013, Pappadopulo:2012}. While the former is specific to 2D and offers optimal OPE convergence properties \cite{Perlmutter:2015, Maldacena:2017, Fitzpatrick:2017, Chang:2019}, the latter generalizes to higher dimensions and has been extensively exploited within the axiomatic approach to CFTs \cite{Kravchuk:2020, Kravchuk:2021, Qiao:2022}. However, these frameworks are analytically intractable for explicit momentum-space calculations. We argue that the global block expansion \cite{Dolan:2000} offers a valuable alternative, as it allows us to recover the time-ordered correlators necessary to construct the causal response function while preserving a mathematically tractable analytic structure. 

Finally, it is worth noting that the momentum-space representations of CFT response functions arise naturally within the AdS/CFT correspondence. In the framework of linear response theory, the poles of these response functions in momentum space determine the characteristic time scale over which a perturbed system relaxes back to equilibrium. These poles were originally analyzed in the context of holographic open quantum systems in~\cite{Loganayagam:2023}. Remarkably, they coincide with the quasinormal modes of the dual black hole, which instead govern the decay of gravitational perturbations around the black hole horizon~\cite{Horowitz:2000, Birmingham:2002}. Consequently, the momentum-space expression of thermal CFT response functions encodes fine-grained information regarding the real-time dynamics of the dual black hole. 

The plan of the paper is the following. In Section $\ref{sec:open}$ we start with the discussion of the dynamics of the open CFT system, by deriving the corrections to its effective dynamics in momentum space. Such terms involve the thermal two-, three- and four-point causal response functions of the environment CFT operators, which are extensively discussed in Section $\ref{sec:correlators}$. In particular, in Sections $\ref{sec:two_point}$ and $\ref{sec:three_point}$ we perform the analytic continuation and the computation of the Fourier transform of the two- and three-point response functions, respectively. The non-trivial analytic structure of the thermal four-point function is investigated in Section $\ref{sec:problem}$, where distinct representations for the conformal blocks are compared. Because the global conformal block expansion is the only representation that remains analytically tractable in momentum space, we also discuss the precise domain in which the real-time continuation of these global blocks can be executed. In Section $\ref{sec:spectral}$ we carry out the computation of the momentum-space expression for the thermal four-point response function, outlining the salient features of the computation. Finally, in Section $\ref{sec:example}$ we specialize our results to the case of a thermal bath modeled by the Ising model.

\section{Effective dynamics of an open quantum system} \label{sec:open}

To motivate the discussion of the Fourier transform of the thermal four-point causal response function of an arbitrary 2D CFT, it is useful to consider a possible scenario where such objects naturally arise, namely the perturbative evaluation of the influence functional associated with the dynamics of an open CFT. 

Let us consider an open quantum system in which both the probe system and the environment are described by two different 2D CFTs in a spacetime with Lorentzian signature. In general, the action of the composite system is
\begin{equation}
    S[\phi, \chi] = S_1 [\phi] + S_2[\chi] + S_{{\text{int}}}[\phi, \chi],
\end{equation}
where $S_1[\phi]$ and $S_2[\chi]$ describe the dynamics of the probe system and environment, respectively. For simplicity, we consider both $\phi$ and $\chi$ to be scalar fields, but nothing forbids us from considering fields with non-zero spin. Without loss of generality, the interaction term can be written as
\begin{equation}
     S_{{\text{int}}}[\phi, \chi] = g \int d^2x \, \mathcal{O}(\phi) \mathcal{V}(\chi),
\end{equation}
where the operators $\mathcal{O}$ and $\mathcal{V}$ are built out of $\phi$ and $\chi$ respectively, but are otherwise arbitrary. The coupling constant $g$ may be needed for dimensional reasons.

\begin{figure}
\centering
\includegraphics[scale=0.2]{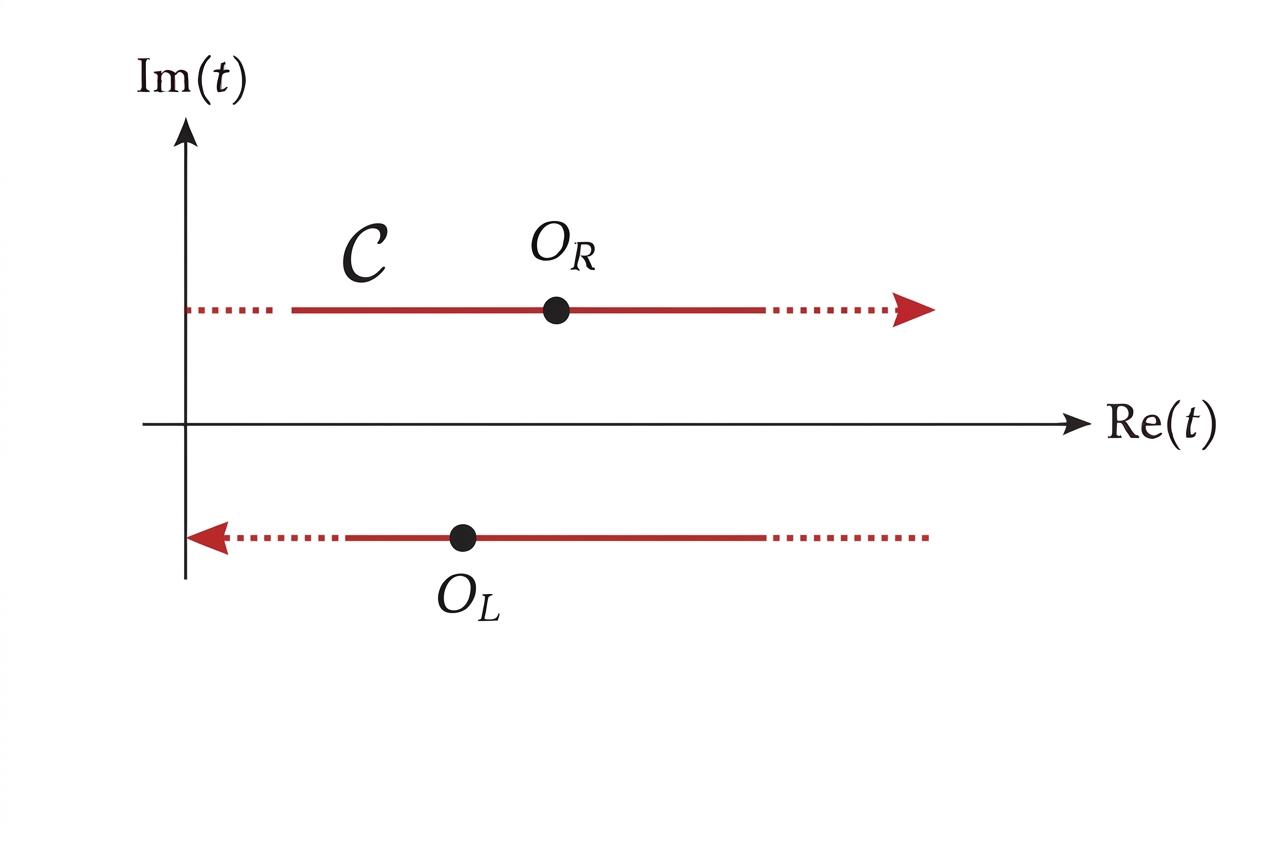}
\vspace*{-15mm}\caption{Illustration of the generic Schwinger-Keldysh complex time contour $\mathcal{C}$, showing the doubling of an arbitrary operator $O$. The upper (forward) and lower (backward) branches allow for the path-integral formulation of real-time dynamics.}
\label{sk_imm}
\end{figure}

Ultimately, we would like to consider the real-time dynamics of the composite system at finite temperature. The most general framework to study a thermal field theory is the Schwinger-Keldysh (SK) formalism~\cite{Haehl:2017, Haehl:2017_2}, where the non-equilibrium dynamics of the system is incorporated through a doubling of the degrees of freedom:
\begin{equation}\label{eq:rl_basis}
    \phi \rightarrow \phi_R, \, \phi_L , \qquad \chi \rightarrow \chi_R, \, \chi_L.
\end{equation}
This doubling reflects the structure of time evolution along a closed contour $\mathcal{C}$ in the complex time plane. As schematically depicted in Figure \ref{sk_imm}, the system evolves forward in time and then backward, ensuring a consistent treatment of expectation values in a density matrix. As a result, causality and unitarity are manifest at the level of the formalism. Thermal initial conditions can be implemented either by an appropriate choice of time contour or, equivalently, through the imposition of the KMS relations~\cite{Haehl:2017_kms}.

In this approach, the classical action becomes 
\begin{equation}
    S = S_1[\phi_L, \phi_R] + S_2[\chi_L, \chi_R] + S_{\text{int}}[\phi_L, \phi_R, \chi_L, \chi_R],
\end{equation}
with 
\begin{equation}\label{eq:int_action}
    \begin{split}
        S_1[\phi_L, \phi_R] = & S_1[\phi_R] - S_1[\phi_L], \qquad S_2[\chi_L, \chi_R] = S_2[\chi_R] - S_2[\chi_L], \\[5mm]
        & S_{\text{int}} = g \int d^2x \, \left( \mathcal{O} (\phi_R) \mathcal{V} (\chi_R)  - \mathcal{O} (\phi_L) \mathcal{V} (\chi_L) \right),
    \end{split}
\end{equation}
whereas the quantum dynamics of the whole system is encoded in the partition function
\begin{equation}
    Z = \int \mathcal{D} \left[ \phi_L, \phi_R, \chi_L, \chi_R \right] e^ {i( S_1[\phi_L, \phi_R] + S_2[\chi_L, \chi_R] + S_{{\text{int}}}[\phi_L, \phi_R, \chi_L, \chi_R])}.
\end{equation}
Since we don't have any control over the degrees of freedom of the environment, the goal is to obtain an effective action describing the dynamics of the degrees of freedom of the probe system alone~\cite{Feynman:1963}. The general idea is to integrate out the field $\chi$ from the partition function, its contribution to the dynamics of the probe system being encoded in the infinite tower of interaction terms contained in the effective action for the probe system:
\begin{equation}
     Z = \int \mathcal{D} \left[ \phi_L, \phi_R \right] e^ {i( S_1[\phi_L, \phi_R] + W[\mathcal{O}(\phi_L, \phi_R)])}.
\end{equation}
The term $W[\mathcal{O}(\phi)]$ is called influence functional and takes the following form:
\begin{equation}
    e^ {i W[\mathcal{O}(\phi_L, \phi_R)]} = \int \mathcal{D} \left[ \chi_L, \chi_R \right] e^ { i( S_2[\chi_L, \chi_R] + S_{{\text{int}}}[\phi_L, \phi_R, \chi_L, \chi_R])}.
\end{equation}
Hence, the problem concerning the dynamics of the probe system boils down to the computation of the influence functional, which contains all the information of the environment and how it affects the system it is coupled with. In the language of standard quantum field theory, $W[\mathcal{O}]$ is the generating functional of the connected diagrams for the environmental field $\chi$. Since the path integral on the right hand side depends on the field $\phi$ only through the interaction term, the influence functional can also be interpreted as a generating functional for the environmental operator $\mathcal{V}$ coupled to a classical current $\mathcal{O}$~\cite{Chandan:2020, Pelliconi:2024}. 

In general, the influence functional cannot be computed exactly. Within the framework of perturbation theory, we can expand $W[\mathcal{O}]$ as a power series in the coupling constant
\begin{equation} \label{eq:infl_funct} 
    i \, W[\mathcal{O}] = \sum_{n=2}^{\infty} \frac{(ig)^{n}}{n!} \int \, d^2x_1 \, ... \int \, d^2x_n \, \langle \mathcal{T}_{\mathcal{C}} \prod_{i=1}^n \left( \mathcal{O}_R(x_i) \mathcal{V}_R(x_i) - \mathcal{O}_L(x_i) \mathcal{V}_L(x_i) \right) \rangle_c,
\end{equation}
where $\mathcal{T}_{\mathcal{C}}$ is the time-ordering operator along the contour in the complex time plane (operators are time-ordered along the forward branch and anti-time ordered along the backward branch) and $\langle ... \rangle_c$ stands for connected diagrams only. Note that the shorthand notation $\mathcal{O}(\phi_R) \equiv \mathcal{O}_R$, $\mathcal{O}(\phi_L) \equiv \mathcal{O}_L$ (and analogous expressions for $\chi_R$ and $\chi_L$) has been introduced. Moreover, the above sum runs from $n=2$ because we assume the one-point functions $\langle \mathcal{V}_R \rangle_c$ and $\langle \mathcal{V}_L \rangle_c$ to vanish. Each term appearing in the expression for $W[\mathcal{O}]$ is nothing but a convolution of a product of operators of the probe system $\mathcal{O}$ at different positions with a correlation function of the environmental operators $\mathcal{V}$. Because of its intrinsic non-local nature, in the most general case the influence functional induces a non-Markovian evolution of the probe system degrees of freedom~\cite{Breuer:2002, Reyes:2026}.

So far the discussion has been general, valid both for equilibrium and non-equilibrium dynamics. Ultimately, our goal is to examine the causal response of the probe system initially in thermal equilibrium to an external disturbance. A remarkable fact about thermal states is that the real-time response functions are related to fluctuations about equilibrium thanks to the fluctuation-dissipation relations, which in turn follow from the KMS relations. This allowed~\cite{Haehl:2017_kms} to construct a causal basis for the full space of $n-$point correlators by singling out from the full set of nested commutators and anticommutators a subset made by only nested commutators where the future-most operator is held fixed. In the single-copy notation this basis looks
\begin{equation}\label{causal_basis}
    \mathcal{B}= \Bigg\{ \langle \left[ \left[...\left[\left[O\left(x_1\right),O\left(x_{\sigma(2)}\right) \right],O\left(x_{\sigma(3)}\right)\right],...\right],O\left(x_{\sigma(n)}\right)\right] \rangle \Bigg\}_{\sigma \in S_{n-1}},
\end{equation}
where $O$ is a generic operator and $S_{n}$ is the set of permutations of $n$ elements. This basis naturally incorporates the restrictions imposed by the KMS conditions and  captures the purely causal response of the system. The response functions defined in standard QFT can be recovered as linear combinations of the elements of the above basis multiplied by the appropriate step functions.

To see how this basis becomes relevant to the previous discussion of the influence functional, we need to express the degrees of freedom of both the probe system and the environment in a different representation, specific to thermal states, called the advanced/retarded basis. The transformation from the R/L representation in ($\ref{eq:rl_basis}$) to the retarded/advanced one takes a particularly simple form in momentum space~\cite{Haehl:2017}. By defining the Fourier transform of the probe field as
\begin{equation}
    \phi(x) = \int \frac{d^2p}{(2 \pi)^2} \, \phi(p) \, e^{-i p x},
\end{equation}
the basis transformation looks 
\begin{equation}\label{eq:ret_adv_basis}
    \begin{split}
        \phi_{adv} (p) = \phi_R (p) - \phi_L (p) , \qquad \phi_{ret} (p) = (f_p + 1)\phi_R (p) - f_p \phi_L (p).
    \end{split}
\end{equation}
Analogous expressions hold for the environmental field $\chi$. Here the function $f(p)$ is the Bose-Einstein distribution function
\begin{equation}
    f_p = \frac{1}{e^{\beta \, p^0}-1},
\end{equation}
which encodes the correct thermal statistics. In this particular representation the interaction action (\ref{eq:int_action}) takes the simple form
\begin{equation} \label{eq:int_action_mom_space}
    S_{\text{int}} = g \int d^2p \, \Big( \mathcal{O}_{adv}(p) \mathcal{V}_{ret}(-p)  + \mathcal{O}_{ret}(p)\mathcal{V}_{adv}(-p) \Big).
\end{equation}
Since the retarded field isolates the component responsible for causal response, while the advanced combination organizes the complementary thermal structure, this representation allows us to easily make contact with the causal response functions involving the environmental field $\chi$ defined in standard QFT. Specifically, the two-, three- and four-point response functions are \cite{Haehl:2017, Haehl:2017_kms}
\begin{equation}\label{eq:response_function}
    \begin{split}
        G_{ra} (x_1, x_2) & = \langle \mathcal{T}_{\mathcal{C}} (\mathcal{V}_{ret}(x_1) \mathcal{V}_{adv}(x_2)) \rangle =  \theta_{12} \langle \left[ \mathcal{V}(x_1), \mathcal{V}(x_2) \right] \rangle, \\[5mm]
        G_{raa}  (x_1, x_2, x_3) & = \langle \mathcal{T}_{\mathcal{C}} \left( \mathcal{V}_{ret} (x_1) \mathcal{V}_{adv} (x_2) \mathcal{V}_{adv} (x_3) \right) \rangle \\
        & = \sum_{\sigma \in S_2} \theta_{1 \sigma(2) \sigma(3)} \langle \left[ \left[\mathcal{V}\left(x_1\right), \mathcal{V}\left(x_{\sigma(2)}\right)\right],\mathcal{V}\left(x_{\sigma(3)}\right)\right] \rangle, \\[3mm]
        G_{raaa} (x_1,x_2,x_3,x_4) & = \langle \mathcal{T}_{\mathcal{C}} (\mathcal{V}_{ret} (x_1) \mathcal{V}_{adv} (x_2)\mathcal{V}_{adv} (x_3) \mathcal{V}_{adv} (x_4)) \rangle \\
        =  \sum_{\sigma \in S_3} \theta_{1 \sigma(2) \sigma(3) \sigma(4)} & \langle \left[ \left[ \left [\mathcal{V} \left( x_1 \right), \mathcal{V} \left( x_{\sigma(2)} \right) \right], \mathcal{V}\left( x_{\sigma(3)} \right) \right], \mathcal{V} \left( x_{\sigma(4)} \right) \right] \rangle,
    \end{split}
\end{equation}
where $\theta_{ij}$ is the usual step function in the time coordinates and $\theta_{ijk} \equiv \theta_{ij} \, \theta_{jk}$. The second equality in the above expressions rewrites the corresponding response function in terms of the elements of the causal basis ($\ref{causal_basis}$). Note that $G_{ra}$ is, up to a factor of $i$, the usual retarded Green's function. 

By using the expressions for the interaction action $(\ref{eq:int_action_mom_space})$ and the response function ($\ref{eq:response_function}$) in the retarded/advanced basis, the perturbative expansion of the influence functional $(\ref{eq:infl_funct})$ takes the form
\begin{equation} \label{eq:final}
    \begin{split}
        i \, W^{(2)} [\mathcal{O}] & = \frac{g^2}{2} \int  d^2 p \, \Big( \mathcal{O}_{adv}(-p) \, \mathcal{O}_{ret}(p) \, \tilde{G}_{ra}(p) \,  + \, ... \Big) ,\\[3mm]
        i \, W^{(3)} [\mathcal{O}] & = - \frac{ig^3}{3!} \int d^2 p \, d^2 p' \, \Big(\mathcal{O}_{adv}(-p) \, \mathcal{O}_{ret}(-p') \mathcal{O}_{ret}(p+p') \, \tilde{G}_{raa}(p,p') \,  + \, ... \Big) , \\[3mm]
        i \, W^{(4)} [\mathcal{O}] & = - \frac{g^4}{4!} \int d^2 p \, d^2 p' \, d^2 p'' \, \Big( \mathcal{O}_{adv}(-p)\,  \mathcal{O}_{ret}(-p') \mathcal{O}_{ret}(-p'') \\
        & \qquad \qquad \qquad \qquad  \qquad \quad \mathcal{O}_{ret}(p+p'+p'') \, \tilde{G}_{raaa}(p,p',p'') \,  + \, ...\Big) ,
    \end{split}
\end{equation}
where $\tilde{G}$ refers to the Fourier transform of $G$. The ellipses stand for all the non-causal terms appearing in the influence functional. As can be seen from these expressions, working in momentum space naturally reduces the degree of non-locality, as convolution structures in position space become algebraic products. In particular, the contribution associated with the two-point response function becomes completely local in momentum space, unlike in position space where it remains non-local. More importantly, the retarded/advanced basis makes the causal structure manifest as many terms that do not contribute to physical response functions are automatically eliminated. 

Since we are considering both the probe system and the environment to be 2D CFTs, the explicit expressions for the thermal two-, three- and four-point functions that construct the response functions appearing in ($\ref{eq:final}$) are well established. Moreover, the corresponding momentum-space representations are also well-known~\cite{Son:2002, Becker:2014}, from which expressions for $\tilde{G}_{ra}$ and $\Tilde{G}_{raa}$ can be directly derived. In the following, we address the remaining gap by computing the Fourier transform of $\tilde{G}_{raaa}$.

\section{Lorentzian correlators}\label{sec:correlators}

In the previous section we ended with an expression for the influence functional involving the Fourier transform of the two-, three- and four-point thermal response functions, valid for any Lorentzian 2D CFT. These functions can be recovered by analytic continuation of the relevant correlators in Euclidean signature, whose form is completely constrained by conformal invariance. In the next section we consider the case of the two-point function, leaving the more complicated discussion of the three-point and four-point functions to subsequent sections.

\subsection{Two-point function}\label{sec:two_point}

To set the stage, let us consider a 2D CFT defined on the Riemann sphere $\hat{\mathbb{C}} = \mathbb{C} \cup \left\lbrace \infty \right\rbrace$ parameterized by the complex coordinates $(z, \bar{z})$. At zero temperature, the two-point function of identical scalar primary fields $\phi(z, \Bar{z})$ with conformal weight $h_\phi$ is defined as
\begin{equation}
    S^{(2)}((z_1,\bar{z}_1), (z_2,\bar{z}_2)) := \langle \phi(z_1, \bar{z}_1) \phi(z_2, \bar{z}_2) \rangle.
\end{equation}
In Euclidean signature, the variables $\Bar{z}_i$ are required to be complex conjugates of $z_i$, $z_i = \Bar{z}_i^*$. This Schwinger function is known to be an analytic function of the variables $z_i \in \mathbb{\hat{C}}$ everywhere except at coincident points $z_i = z_j$ ($i \neq j$). Conformal invariance constrains the two-point function to take the form
\begin{equation} 
        S^{(2)} (z_1, z_2) =  \frac{1}{z_{12}^{2 h_\phi} \bar{z}_{12}^{2 h_\phi}}, \qquad \qquad z_{ij} = z_i - z_j.
\end{equation}
Translation invariance also allows us to set one coordinate equal to zero, usually $z_2=\bar{z}_2=0$.

The corresponding CFT at finite temperature $T$ can be obtained by conformally mapping the complex plane $z=x + i \tau $ to a cylinder whose time coordinate has period $\beta = 1/T$: 
\begin{equation}
    z \to e^{2 \pi T z}, \qquad  \tau \sim \tau + \beta.
\end{equation}
The two-point function then becomes~\cite{Son:2002}
\begin{equation}\label{remark}
        S^{(2)}_\beta (z_1, z_2) = \frac{(\pi T)^{2 h_\phi}}{\sinh^{2 h_\phi}(\pi T z_{12})} \frac{(\pi T)^{2 h_\phi}}{\sinh^{2 h_\phi}(\pi T \bar{z}_{12})}.
\end{equation}

So far, the discussion has been in Euclidean signature. Lorentzian correlators can be recovered from their Euclidean counterparts by analytic continuation. For the sake of clarity, we are going to label points on the Riemann sphere as $z=(\tau,x) \in \mathbb{\hat{C}}$, while those in the Minkowski spacetime as $y=(t,x) \in \mathbb{R}^{1,1}$. Without loss of generality, we fix the (Euclidean) time ordering $\tau_1  >...> \tau_n$.

In general, the analytic continuation of any $n$-point Schwinger function to a corresponding $n$-point Wightman function is performed along a continuous path in configuration space that connects the initial Euclidean configuration to the final Lorentzian one. In practice, this is implemented by relaxing the condition $z_i = \bar{z}_i$ and performing a Wick rotation of each Euclidean time coordinate~\cite{Haang:1992, Hartman:2019}
\begin{equation}
    \tau_k =  i t_k - \epsilon_k, \qquad \epsilon_k \in \mathbb{R}^{+}, \qquad k=1,...,n,
\end{equation}
where each Euclidean time is allowed to have an infinitesimal constant real part. Different relative orderings of the $\epsilon_k$ parameters prescribe distinct paths in the complex $\tau$-plane along which the analytic continuation is executed. These paths wrap around the light-cone branch cuts in non-equivalent ways, thereby generating the various permutations of causal orderings for the resulting Lorentzian correlators. Whenever a light-cone branch cut is crossed, the continuation path transitions onto a higher Riemann sheet of the complex domain. Ultimately, each individual $n$-point Wightman function is recovered by taking the limit $\epsilon_k \to 0$ while strictly preserving the relative ordering:
\begin{equation}
    W^{(n)} (y_{\sigma(1)},...,y_{\sigma(n)}) = \lim_{\substack{\epsilon_1,...,\epsilon_n \rightarrow 0 \\ \epsilon_{\sigma(1)}<...<\epsilon_{\sigma(n)}}} S^{(n)} ((x_1,\tau_1), ..., (x_n,\tau_n))\Big|_{\tau_i= i t_i - \epsilon_i}, \qquad \sigma \in S_n.
\end{equation}

The analytic continuation of the thermal two-point function is almost trivial, as it follows from a blind application of the above $i \epsilon-$prescription~\cite{Son:2002, Becker:2014}
\begin{equation}
    \begin{split}
        W^{(2)}_\beta (y_1, y_2) & = S^{(2)}_\beta((x_1, \tau=i t_1 - \epsilon_1), (x_2, \tau_2 = i t_2 - \epsilon_2)) \\
        & =e^{- 2 \pi i h_\phi} \frac{(\pi T )^{2 h_\phi}}{\sinh^{2 h_\phi}(\pi T (x_1^-- x_2^-))} \frac{(\pi T )^{2 h_\phi}}{\sinh^{2 h_\phi}(\pi T (x_1^+ - x_2^+))}, 
    \end{split}
\end{equation}
where we introduce the light-cone coordinates $x_i^\pm = t_i \pm x_i$ $(i=1,2)$ and we fix the ordering $\epsilon_2 > \epsilon_1$. Hence, the only relevant effect is encoded in the overall phase factor.

As outlined in the previous section, within the framework of open quantum systems we are more interested in the causal response functions rather than the above Wightman function. Upon analytic continuation, the two-point response function $G_{ra}$ in (\ref{eq:response_function}) takes the form
\begin{equation}
    \begin{split}
        G_{ra}(y_1,y_2) &= \theta_{12} \left( W^{(2)}_{\beta}(y_1, y_2) - W^{(2)}_{\beta} (y_2, y_1) \right) \\
        & = - 2 i \sin(2 \pi h_\phi) \, \theta_{12} \, \frac{(\pi T )^{2 h_\phi}}{\sinh^{2 h_\phi}(\pi T (x_1^--x_2^-))} \frac{(\pi T )^{2 h_\phi}}{\sinh^{2 h_\phi}(\pi T (x_1^+-x_2^+))}.
    \end{split}
\end{equation}
By defining the spectral function\footnote{A note on nomenclature: although the expression \textquotedblleft spectral function" originally denoted the imaginary part of the retarded propagator, it is nowadays also commonly used to indicate the Fourier transform of the corresponding commutator.} $\rho_{ra}(p_1,p_2)$ as the Fourier transform of the commutator\footnote{The definition of spectral function in ($\ref{eq:def_spect_fct}$) usually includes an extra factor of $-i$. However, our definition is consistent with the expressions for the response functions given in ($\ref{eq:response_function}$).}
\begin{equation}\label{eq:def_spect_fct}
    \rho_{ra}(p_1,p_2) =  \int \, d^2 y_1 \, e^{i p_1 y_1} \int \, d^2 y_2 \, e^{i p_2 y_2} \,  \langle [\phi(y_1),\phi(y_2)] \rangle,
\end{equation}
the response function can be recast into~\cite{Chaudhuri:2019}
\begin{equation}
    G_{ra} (y_1,y_2) = \int \frac{d^2p_1}{(2 \pi)^2} \, e^{-i p_1 y_1} \int \frac{d^2p_2}{(2 \pi)^2}  \, e^{-i p_2 y_2} \, \theta_{12} \, \rho_{ra}(p_1,p_2).
\end{equation}
The momentum-space expression of $G_{ra}$ is then fixed in terms of the spectral function $\rho_{ra}$. By setting $y_2=0$ and writing the momentum coordinate as $p=(\omega, k)$, the momentum-space expression of the two-point spectral function simplifies to
\begin{equation} \label{integral_1}
    \begin{split}
        \rho_{ra}(p) = - i \sin(2 \pi h_\phi) \frac{(2 \pi T)^{2( 2 h_\phi -1)}}{\Gamma^2 \left(2 h_\phi \right)} \prod_{\delta=\pm} \Gamma \left( h_\phi + \frac{i p^\delta}{2 \pi T} \right) \Gamma \left( h_\phi - \frac{i p^\delta}{2 \pi T} \right),
    \end{split}
\end{equation}
where we introduce the light-cone momentum coordinates $p^{\pm}= \frac{- \omega \pm k}{2}$. This result follows from the basic integral~\cite{Gubser:1997, Son:2002, Manenti:2020}
\begin{equation} \label{basic_integral}
    \int dx \, e^{ipx} \, \frac{(\pi T)^\Delta}{\sinh^\Delta(\pi T x)} = \frac{(2 \pi T)^{\Delta-1}}{\Gamma(\Delta)} \, \Gamma \left( \frac{\Delta}{2} + \frac{ip}{2 \pi T} \right) \, \Gamma \left( \frac{\Delta}{2} - \frac{ip}{2 \pi T} \right),
\end{equation}
which is convergent only in the interval $0 < \Delta <1$. However, since the result $(\ref{integral_1})$ is analytic in the parameter $h_\phi$, it can be analytically continued to any value of the complex $h_\phi-$plane such that
\begin{equation}
    h_\phi > 0, \qquad - h_\phi \pm \frac{i p^+}{2 \pi T} \notin \mathbb{N}^0, \qquad - h_\phi \pm \frac{i p^-}{2 \pi T} \notin \mathbb{N}^0.
\end{equation} 
While the first condition is the usual unitary bound for the conformal weight of any CFT field, the remaining conditions account for the poles of the Gamma functions, which perfectly match the quasinormal modes of the dual BTZ black hole~\cite{Birmingham:2002}.

\subsection{Three-point function}\label{sec:three_point}

Let's now consider the case of the three-point function of identical primary scalar fields $\phi(z, \bar{z})$. Conformal invariance fixes the form of the three-point function up to an overall constant $C (h_\phi)$:
\begin{equation}
    S^{(3)}(z_1, z_2, z_3) =  C (h_\phi) \frac{1}{z_{12}^{h_\phi} z_{23}^{h_\phi} z_{13}^{h_\phi}} \frac{1}{\bar{z}_{12}^{h_\phi} \bar{z}_{23}^{h_\phi} \bar{z}_{13}^{h_\phi}}.
\end{equation}
Analogously to the case of the two-point function, the above Schwinger function is an analytic function of the variables $z_i \in \hat{\mathbb{C}}$ everywhere except at coincident points $z_i = z_j$ ($i \neq j$). Performing the usual conformal map from the complex plane to the cylinder yields the three-point function at finite temperature~\cite{Becker:2014}:
\begin{equation}
    \begin{split}
        S^{(3)}_\beta (z_1, z_2, z_3) = C (h_\phi) \prod_{i<j} \frac{(\pi T)^{h_\phi}}{\sinh^{h_\phi}(\pi T z_{ij})} \frac{(\pi T)^{h_\phi}}{\sinh^{h_\phi}(\pi T \bar{z}_{ij})}.
    \end{split}
\end{equation}

Analytic continuation of the thermal three-point function proceeds in a straightforward manner by applying the $i\epsilon$-prescription outlined in the previous section. Because the deformation paths remain entirely on the principal Riemann sheet, the structural form of the correlator is preserved, with the analytic continuation manifesting simply as an overall phase factor:
\begin{equation}
    W^{(3)}_\beta(y_1, y_2, y_3) = e^{-3 \pi i h_\phi} C (h_\phi) \prod_{i<j} \prod_{\delta= \pm} \frac{(\pi T)^{h_\phi}}{\sinh^{h_\phi}(\pi T (x_i^\delta-x_j^\delta))}.
\end{equation}
Here we fixed the ordering $\epsilon_1 < \epsilon_2 < \epsilon_3$ and we switched to the more compact notation associated with the light-cone coordinates $x_i^\pm = t_i \pm x_i$.

As discussed in Section $\ref{sec:open}$, the relevant three-point response function in (\ref{eq:response_function}) involves a linear combination of commutators multiplied by the relevant step functions. Performing the analytic continuation of such commutators yields
\begin{equation}
    \begin{split}
        G_{raa} (y_1, y_2, y_3) =&  - 4 \, \sin(2 \pi h_\phi) \, \sin (\pi h_\phi) \, C (h_\phi) \\
        &\Bigg( \theta_{123} \prod_{\delta= \pm} \prod_{i<j} \frac{(\pi T)^{h_\phi}}{\sinh^{h_\phi}(\pi T (x_i^\delta-x_j^\delta))} + (2 \leftrightarrow 3) \Bigg).
    \end{split}
\end{equation}
By setting $y_3=0$, the above expression can be greatly simplified. In particular, this allows us to define the momentum-space expression of $G_{raa}$ in terms of a single three-point spectral function $\rho_{raa}(p_1, p_2)$:
\begin{equation}
    G_{raa}(y_1,y_2) = \int \frac{d^2 p_1}{(2 \pi)^2} \, e^{- i p_1 y_1} \int \frac{d^2 p_2}{(2 \pi)^2} \, \, e^{-i p_2 y_2} \left(\theta_{123} + (-1)^{2 h_\phi}\theta_{132} \right) \, \rho_{raa}(p_1, p_2).
\end{equation}
The expression of the three-point spectral function can still be determined analytically~\cite{Becker:2014}. Upon defining the momentum-space light-cone coordinates $p_i^{\pm} = \frac{- \omega_i \pm k_i}{2}$, the spectral function factorizes as
\begin{equation}
    \begin{split}
        \rho_{raa} (p_1, p_2) =  - \sin(2 \pi h_\phi) \, \sin (\pi h_\phi) \, C (h_\phi) \, \prod_{\delta= \pm} I (p_1^\delta, p_2^\delta),
    \end{split}
\end{equation}
with
\begin{equation}
    \begin{split}
        I(p_1,p_2) = \int \left( \prod_{j=1}^2 d x_j \, e^{i p_j x_j} \right)  \frac{(\pi T )^{h_\phi}}{\sinh^{h_\phi} (\pi T x_1)} \frac{(\pi T )^{h_\phi}}{\sinh^{h_\phi} (\pi T x_2)} \frac{(\pi T )^{h_\phi}}{\sinh^{h_\phi} (\pi T (x_1 - x_2))}.
    \end{split}
\end{equation}
By defining $u = x_1 - x_2$ and inserting a delta function
\begin{equation}
    \delta(u - x_1 + x_2) = \int \frac{d \omega}{2 \pi} \, e^{- i \omega (u - x_1 + x_2)}
\end{equation}
the integral $I(p_1,p_2)$ can be factorized into
\begin{equation}
\begin{split}
        I(p_1,p_2) = \int \frac{d \omega}{2 \pi} \int  d x_1 \frac{(\pi T)^{ h_\phi} e^{ i ( p_1 + \omega) x_1}}{\sinh^{h_\phi} (\pi T x_1)} \int d x_2 \frac{(\pi T)^{ h_\phi} e^{i (p_2 - \omega) x_2}}{\sinh^{h_\phi} (\pi T x_2)} \int d u \frac{(\pi T)^{ h_\phi} e^{- i \omega u}}{\sinh^{h_\phi} (\pi T u)} .
    \end{split}
\end{equation}
Exploiting the basic integral in ($\ref{basic_integral}$) yields the following expression for the three-point spectral function
\begin{equation}
    \begin{split} \label{3_pt_spctr_lor}
        \rho_{raa} (p_1,p_2) = & -\sin(2 \pi h_\phi) \, \sin (\pi h_\phi) \, C (h_\phi) \, \frac{(2 \pi T)^{2(3 h_\phi -2)}}{\Gamma^6 \left( h_\phi \right)} \\
        & \prod_{\delta = \pm} G_{3,3}^{3,3} 
         \begin{pmatrix}
            1 - \frac{h_\phi}{2} - \frac{i p_2^\delta}{2 \pi T}, & 1 -\frac{h_\phi}{2}, & 1 -  \frac{h_\phi}{2} + \frac{i p_1^\delta}{2 \pi T} \\ 
            & & & & e^{i \pi} \\
             \frac{h_\phi}{2} - \frac{i p_2^\delta}{2 \pi T}, &  \frac{h_\phi}{2}, &  \frac{h_\phi}{2} + \frac{i p_1^\delta}{2 \pi T} 
        \end{pmatrix},
    \end{split}
\end{equation}
where $G_{3,3}^{3,3}$ represents the Meijer G-function~\cite{Gfunction}. For a generic Meijer G-function
\begin{equation}
    G^{\,m,n}_{\,p,q} 
    \begin{pmatrix}
            a_1, ..., a_p \\
            & z \\
             b_1,...,b_q 
        \end{pmatrix}, \qquad 0 \leq n \leq p, \, 0 \leq m \leq q,
\end{equation}
the regularity conditions
\begin{equation}\label{eq:poles_cond}
    \begin{split}
        a_k - b_j & \notin \mathbb{N}, \qquad k = 1, ..., n, \quad j = 1,..., m,  \\
        b_i - b_j & \notin \mathbb{Z}, \qquad i, j= 1,..., m, \quad i \neq j,
    \end{split}
\end{equation}
ensure the existence of the underlying Mellin-Barnes contour integral. These constraints also allow us to read off the pole structure of the above spectral function
\begin{equation}
    \begin{split}
        &\frac{i p_1^\delta}{2 \pi T} \notin  \mathbb{Z}, \qquad \qquad \qquad \frac{i p_2^\delta}{2 \pi T} \notin \mathbb{Z}, \qquad \qquad \qquad \frac{i (p_1^\delta + p_2^\delta)}{2 \pi T} \notin \mathbb{Z}, \\
        &1 - h_\phi \pm \frac{i p_1^\delta}{2 \pi T} \notin \mathbb{N}, \qquad 1 - h_\phi \pm \frac{i p_2^\delta}{2 \pi T} \notin \mathbb{N}, \qquad  1 - h_\phi \pm \frac{i (p_1^\delta + p_2^\delta)}{2 \pi T} \notin \mathbb{N},
    \end{split}
\end{equation}
with $\delta = \pm$. Again, these poles correspond to the quasinormal modes of the dual BTZ black hole. Our result differs from the original holographic computation~\cite{Loganayagam:2023} due to a different choice of basis for the thermal response functions. Specifically, the authors of~\cite{Loganayagam:2023} adopted an elegant basis introduced in~\cite{Chaudhuri:2019} to study thermal OTO correlators. This basis is just a variant of the retarded/advanced representation (\ref{eq:ret_adv_basis}), where thermal statistical factors arising from the KMS constraints are systematically factored out of the real-time correlators. Consequently, the three-point response function discussed in~\cite{Loganayagam:2023} differs from $G_{raa}$ because it comprises a different linear combination of the elements of the causal basis ($\ref{causal_basis}$). Ultimately, the pole structures of the two computations can be mapped to one another via an appropriate change of basis.

\subsection{Four-point function}\label{sec:four_point}

So far, we have examined the momentum-space representations of the thermal two- and three-point functions, for which the analytic continuation can be readily implemented via a standard application of the $i \epsilon-$prescription. We now turn to the more intricate case of the four-point function, whose non-trivial analytic structure renders analytic continuation significantly more challenging. In the next section, we discuss various kinematic representations of the four-point function, each possessing distinct convergence properties that dictate its behavior under analytic continuation. The explicit momentum-space expression of the thermal four-point response function, which plays a central role in the analysis of the open CFT discussed in Section \ref{sec:open}, is derived in Section \ref{sec:spectral}.

\subsubsection{Kinematic representations}\label{sec:problem}

At zero temperature, the four-point function of identical scalar fields $\phi(z, \bar{z})$ is constrained by conformal invariance to take the form
\begin{equation}\label{eq:4_pt_fct}
        S^{(4)} (z_1, z_2, z_3, z_4) =  \frac{g(z, \bar{z})}{z_{12}^{2h_\phi} z_{34}^{2h_\phi} \bar{z}_{12}^{2h_\phi} \bar{z}_{34}^{2h_\phi}}, 
\end{equation}
where $g(z, \bar{z})$ is a function of the conformally invariant cross ratios $z$ and $\bar{z}$:
\begin{equation}\label{eq:cross_ratio_zero}
    z = \frac{z_{12} z_{34}}{z_{13} z_{24}}, \qquad \bar{z} = \frac{\bar{z}_{12} \bar{z}_{34}}{\bar{z}_{13} \bar{z}_{24}}.
\end{equation}
This Schwinger function is an analytic function of the variables $z_i \in \hat{\mathbb{C}}$ everywhere except at coincident points $z_i = z_j$ ($i \neq j$). By exploiting conformal invariance we can map $\lbrace z_1,z_2,z_3,z_4\rbrace$ to the standard configuration $\lbrace 0,z,1,\infty \rbrace$.

The function $g(z, \bar{z})$ admits several distinct representations. Here we are primarily concerned with its decomposition into global conformal blocks, as this specific framework facilitates the explicit evaluation of the corresponding Fourier transform. By considering the operator combination $(12)(34)$ and applying the OPE twice, the $s$-channel expansion of $g(z, \bar{z})$ looks
\begin{equation}\label{eq:ope_dec}
    g(z, \bar{z}) = \sum_{h, \bar{h}} C^2 (h_\phi; h, \bar{h}) \, f (h,h_\phi;z) \, f (\bar{h}, h_\phi; \bar{z}),
\end{equation}
where the sum is over all possible global primary fields $O_{h, \bar{h}}$ that appear in the $\phi \times \phi$ OPE, and $C (h_\phi; h, \bar{h})$ is the structure constant associated with the three-point function $\langle \phi \phi O_{h, \bar{h}} \rangle$. The global conformal block $f (h,h_\phi;z)$ contains all the contributions to the four-point function related to the descendants of $O_{h, \bar{h}}$. As pointed out in ~\cite{Dolan:2000}, global conformal blocks of any unitary 2D CFT can be expressed in terms of hypergeometric functions:
\begin{equation}\label{eq:s_global_blocks}
    \begin{split}
        f (h,h_\phi;z) = z^h \, {}_2F_1 \left( h, h, 2h, z \right).
    \end{split}
\end{equation}
Each global block is analytic in the region $\mathbb{C} \setminus [1, +\infty)$. While standard radial quantization arguments naively predict that the full OPE series for $g(z,\bar{z})$ converges absolutely only within the open unit disk $|z|<1$, the actual domain of convergence can be extended to the entire cut plane $\mathbb{C} \setminus [1, +\infty)$ \cite{Pappadopulo:2012}. 

Naturally, the $s$-channel expansion is tailored to describe the kinematic regime where $\phi(z_1)$ and $\phi(z_2)$ are close to one another, providing an asymptotic expansion around $z \to 0$. However, nothing forbids us from considering other operator combinations, namely the $t-$channel $(14)(23)$ and the $u-$channel $(13)(24)$. Explicitly, the global block decomposition in these alternative channels is obtained by implementing the following replacements to ($\ref{eq:4_pt_fct}$), ($\ref{eq:cross_ratio_zero}$) and ($\ref{eq:s_global_blocks}$):
\begin{equation}\label{eq:t_u_channel}
    \begin{split}
        & \text{t-channel:} \quad z_2\to z_4, \quad z_3 \to z_2, \quad z_4 \to z_3, \quad z \to  1-z, \\
        & \text{u-channel:} \quad z_2\to z_3, \quad z_3 \to z_2, \quad z \to \frac{1}{z}.
    \end{split}
\end{equation}
Obviously, the $t-$channel takes the form of a series expansion around $z\to1$ convergent in the cut plane $\mathbb{C} \setminus (-\infty,0]$, whereas the $u-$channel is associated with an expansion around $z\to \infty$ convergent in the cut plane $\mathbb{C} \setminus [0,1]$. Crucially, the physical four-point function must satisfy crossing symmetry, viz. the different channel expansions $g(z,\bar{z})$, $g(1-z,1-\bar{z})$, $g(1/z,1/\bar{z})$ must coincide in the regions of the complex $z-$plane where they are jointly valid. These consistency conditions provide a tower of constraints on the CFT data, forming the backbone of the conformal bootstrap approach \cite{Poland:2019, Iliesiu:2018}.

To compute the four-point response function for the open CFT introduced in Section \ref{sec:open}, we must analytically continue the Euclidean four-point correlators to Lorentzian signature. In principle, this requires tracking $4!=24$ different time orderings of the four-point function, 8 of which correspond to OTO correlators. While these OTO correlators have found numerous applications, most notably in the study of quantum chaos \cite{Roberts:2015, Maldacena:2016, Chang:2019}, they do not contribute to our analysis. This is because the physical response function is constructed strictly from causal, time-ordered commutators.

Focusing for definiteness on the $s$-channel expansion of $g(z,\bar{z})$, term-by-term analytic continuation is valid provided that the paths in the $z,\,\bar{z}-$planes do not cross the branch cut $[1,+\infty)$, ensuring $g(z,\bar{z})$ stays on the principal Riemann sheet. Crossing this cut transitions the system onto the second Riemann sheet, where the $s$-channel expansion generally ceases to converge. Causal orderings that force a crossing of the $[1,+\infty)$ cut map to the aforementioned OTO correlators. Conversely, paths that avoid the cut yield the standard time-ordered correlators used in the next section to construct the causal response function.

Notably, while one might hope to circumvent an $s$-channel breakdown by switching to an alternative channel, it is well established that certain causal configurations exist where all channel expansions simultaneously diverge \cite{Qiao:202222}. A prominent example of this behavior occurs in the Regge limit \cite{Costa:2012}, where $z_1$, $z_4$ and $z_2$, $z_3$ are timelike separated while all other intervals are spacelike. This absolute divergence is a primary motivator for exploring alternative conformal block frameworks.

An effective alternative that resolves these convergence limitations is the radial coordinate expansion. This framework leverages a kinematic map that projects the complex cut plane onto the interior of the unit disk. The corresponding change of coordinates is given by \cite{Hogervorst:2013}
\begin{equation}\label{eq:radial}
    \rho = \frac{z}{\left( 1 - \sqrt{1-z} \right)^2}, \qquad \bar{\rho} = \frac{\bar{z}}{\left( 1 - \sqrt{1-\bar{z}} \right)^2},
\end{equation}
such that the conformal block results in a power series in $\rho, \, \bar{\rho}$ (we abuse the notation a bit by writing $f (h,h_\phi;z)$ or $f (h,h_\phi;\rho)$ depending on which set of coordinates we want to use)
\begin{equation}
    f (h,h_\phi;\rho) = \sum_{\substack{n>0, \, m \in 2 \mathbb{Z} \\ |m|<n}} p_{n,m} \, r^n  e^{i\, m \, \theta}, \qquad \rho = r \, e^{i\, \theta}, \qquad p_{n,m} \geq 0 \quad \forall \, \,n,\,m.
\end{equation}
Consequently, the OPE decomposition of $g(\rho, \bar{\rho})$ converges absolutely within the unit disk $|\rho|, \, |\bar{\rho}|<1$, which corresponds to $\mathbb{C} \setminus [1,+\infty)$ in the original $z$-plane. An analogous construction holds for the $t$- and $u$-channels upon replacing $z$ in the transformation law ($\ref{eq:radial}$) according to ($\ref{eq:t_u_channel}$). 

These distinct channel expansions, however, exhibit highly non-equivalent properties upon analytic continuation to Lorentzian signature. Crucially, for the $s$-channel expansion, the entire Lorentzian kinematic domain maps onto the closure of the unit disk $|\rho|, |\bar{\rho}| \leq 1$, a property that does not hold for the other channels \cite{Kravchuk:2020}. Consequently, it is possible to show that $g(\rho, \bar{\rho})$ converges in the sense of distributions also on the boundary of this unit disk. This property ensures that the $s-$channel analytic continuation remains mathematically well-defined for any causal ordering \cite{Kravchuk:2021, Qiao:2022}. 

While the radial-coordinate block representation is highly efficient for higher-dimensional CFTs, it is sub-optimal for 2D CFTs, where the infinite-dimensional Virasoro symmetry can be leveraged. The optimal framework in 2D utilizes the Virasoro conformal blocks, which can be interpreted as a reorganization of the global blocks ($\ref{eq:s_global_blocks}$) into the much larger representations of the full Virasoro symmetry algebra. In this framework, the OPE decomposition is structurally analogous to $(\ref{eq:ope_dec}$), namely
\begin{equation}
    g(q, \bar{q}) = \sum_{h, \bar{h}} C^2 (h_\phi; h, \bar{h}) \, f (h,h_\phi;q) \, f (\bar{h}, h_\phi; \bar{q}),
\end{equation}
with the crucial distinction that the sum now runs strictly over the Virasoro primary fields $\mathcal{V}_{h,\bar{h}}$. The Virasoro block $f (h,h_\phi;q)$ admits the elliptic representation \cite{Zamolodchikov:1987}
\begin{equation}\label{eq:virasoro_block}
    f (h,h_\phi;q) = (16 q)^{h-\frac{c-1}{24}} \, z^{\frac{c-1}{24}} \, (1-z)^{\frac{c-1}{24}-2 h_\phi} \, \vartheta_3(q)^{\frac{c-1}{2} - 16 h_{\phi}} \, H(h;q),
\end{equation}
where $\vartheta_3(q)$ is the Jacobi theta function \cite{Polchinski:2007}, $c$ is the central charge of the CFT, and $q$ is the Zamolodchikov uniformizing variable 
\begin{equation}\label{eq:q}
    q(z) = e^{i \pi \tau(z)}, \qquad \tau(z) = i \, \frac{{}_2F_1 \left( \frac{1}{2}, \frac{1}{2}, 1, 1-z \right)}{{}_2F_1 \left( \frac{1}{2}, \frac{1}{2}, 1, z \right)}.
\end{equation}
The Virasoro block is completely specified up to the function $H(h;q)$, which can be expanded as a power series around $q=0$ with recursively determined coefficients \cite{Perlmutter:2015}. 

Under this elliptic coordinate transformation, the points $\lbrace 0, 1, \infty \rbrace$ in the $z$-plane are mapped to the points $\lbrace 0, 1, -1 \rbrace$ in the $q$-plane. Consequently, the entire $z$-plane is mapped to a compact subregion of the unit circle $|q| \leq 1$. The core advantage of this representation is its absolute convergence within the unit disk $|q|, \, |\bar{q}|<1$, which is strictly larger than the domain $|\rho|, \, |\bar{\rho}| < 1$. This property ensures that the Virasoro OPE decomposition of $g(q,\bar{q})$ remains stable under arbitrary analytic continuations \cite{Maldacena:2017, Fitzpatrick:2017}.

In summary, comparing the various expressions for the conformal block reveals that only the global block representation (\ref{eq:s_global_blocks}) is mathematically tractable in momentum space. In the following section, we recover the time-ordered correlators necessary to construct the thermal four-point response function and analyze its explicit momentum-space representation.

\subsubsection{Four-point spectral function} \label{sec:spectral}

As outlined in the previous section, the only representation fully amenable to evaluating the influence functional of an open CFT in momentum space is the global conformal block expansion ($\ref{eq:s_global_blocks}$). In this representation, the thermal four-point function is given by
\begin{equation} \label{eq:general_4pt}
    \begin{split}
        S^{(4)}_\beta(z_1, z_2, z_3, z_4) = & \frac{(\pi T)^{2 h_\phi}}{\sinh^{2 h_\phi}(\pi T z_{12})} \frac{(\pi T)^{2h_\phi}}{\sinh^{2 h_\phi}(\pi T z_{34})} \frac{(\pi T)^{2 h_\phi}}{\sinh^{2 h_\phi}(\pi T \bar{z}_{12})} \frac{(\pi T)^{2 h_\phi}}{\sinh^{2 h_\phi}(\pi T \bar{z}_{34})} \\
        & \sum_{h, \bar{h}} C^2 (h_\phi; h, \bar{h}) \, z^{h} \, {}_2F_1 (h, h, 2h, z) \, \bar{z}^{\bar{h}} \, {}_2F_1 (\bar{h}, \bar{h}, 2\bar{h},\bar{z}), 
    \end{split}
\end{equation}
where the thermal cross ratios are  
\begin{equation}\label{eq:thermal_cross}
    z = \frac{\sinh(\pi T z_{12}) \sinh(\pi T z_{34})}{\sinh(\pi T z_{13}) \sinh(\pi T z_{24})}, \qquad \bar{z} = \frac{\sinh(\pi T \bar{z}_{12}) \sinh(\pi T \bar{z}_{34})}{\sinh(\pi T \bar{z}_{13}) \sinh(\pi T \bar{z}_{24})}.
\end{equation}
The OPE decomposition is still valid on the cylinder \cite{Pappadopulo:2012}. As previously established in the discussion below ($\ref{eq:t_u_channel}$), a term-by-term analytic continuation of the above correlator can be performed as long as the paths in the $z, \, \bar{z}-$planes do not cross any branch cut. While this constraint precludes the recovery of OTO configurations, it is sufficient for extracting the standard time-ordered correlators necessary to construct the causal response function.

To be more explicit, let us fix the ordering $\epsilon_1 < \epsilon_2 < \epsilon_3 < \epsilon_4$ corresponding to the fully time-ordered correlator. The conformal prefactor in ($\ref{eq:general_4pt}$) can be analytically continued by a straightforward application of the $i \epsilon-$prescription: it simply picks up a phase factor equal to $e^{-4 \pi i h_\phi}$. On the other hand, each global block retains the same structure under analytic continuation, the only modification being that the cross ratios become independent variables:
\begin{equation}\label{cross_ratio_mink}
    \begin{split}
        z & = \frac{\sinh(\pi T (x_1^- - x_2^-))  \sinh(\pi T (x_3^--x_4^-))}{\sinh(\pi T (x_1^- - x_3^-))\sinh(\pi T (x_2^- - x_4^-))}, \\
        \bar{z} & = \frac{\sinh(\pi T (x_1^+ - x_2^+)) \sinh(\pi T (x_3^+ - x_4^+))}{\sinh(\pi T (x_1^+ - x_3^+)) \sinh(\pi T (x_2^+ - x_4^+))}.
    \end{split}
    \end{equation}
Overall, the fully time-ordered four-point correlator takes the form
\begin{equation} \label{eq:thermal_4_lor}
    \begin{split}
        W^{(4)}_\beta (y_1, y_2, y_3, y_4) = e^{-4 \pi i h_\phi} \, w^{(4)}_{\beta}(y_1, y_2, y_4, y_4),
    \end{split}
\end{equation}
where, for later convenience, we define the structure factor
\begin{equation} \label{eq:true_4}
    \begin{split}
        w^{(4)}_{\beta}(y_1, y_2, y_4, y_4) = & \left( \prod_{\delta= \pm} \frac{(\pi T)^{2 h_\phi}}{\sinh^{2 h_\phi}(\pi T (x_1^\delta - x_2^\delta))} \frac{(\pi T)^{2 h_\phi}}{\sinh^{2 h_\phi}(\pi T (x_3^\delta - x_4^\delta))} \right) \\
        & \sum_{h, \bar{h}} C^2 (h_\phi; h, \bar{h}) \, z^{h} \,  {}_2F_1 (h, h, 2h, z) \, \bar{z}^{\bar{h}} \, {}_2F_1 (\bar{h}, \bar{h}, 2\bar{h},\bar{z}).
    \end{split}
\end{equation}

Equipped with this expression for the four-point function in Lorentzian signature, one can construct the four-point response function in $(\ref{eq:response_function})$ by taking appropriate linear combinations of $(\ref{eq:true_4})$ and enforcing the correct causal structure through step functions of time coordinate differences:
\begin{equation}
    \begin{split}
        G_{raaa}(y_1,y_2,y_3,y_4) =   8 i \sin(3 \pi h_\phi) \sin(2 \pi h_\phi) \sin( \pi h_\phi) & \\
        \sum_{\sigma \in S_3} \theta_{1 \sigma(2) \sigma(3) \sigma(4)} \, w^{(4)}_\beta \left(y_1,y_{\sigma(2)},y_{\sigma(3)},y_{\sigma(4)} \right).&
    \end{split}
\end{equation}

Similarly to the previous cases, the four-point spectral function can be defined through the Fourier transform
\begin{equation}
    \begin{split}
        G_{raaa} (y_1,y_2,y_3) = \int \left( \prod_{j=1}^3  \frac{d^2 p_j}{(2 \pi)^2} \, e^{-i p_j y_j} \right)  \Big(  (\theta_{1234} + \theta_{1243} + \theta_{1342})\rho_{raaa}(p_1,p_2,p_3) & \\
        + (\theta_{1423} +\theta_{1432} +\theta_{1324})\rho_{raaa}(p_1,p_3,p_2) \Big)&,
    \end{split}
\end{equation}
from which we can isolate 
\begin{equation}
    \begin{split}
        \rho_{raaa}(p_1, p_2, p_3) = & 8 i \sin(3 \pi h_\phi) \sin(2 \pi h_\phi) \sin( \pi h_\phi) \\
        &\int \left( \prod_{j=1}^3 d^2 y_j \, e^{i p_j y_j} \right) w^{(4)}_\beta \left(y_1,y_2,y_3 \right).
    \end{split}
\end{equation}
Evaluating the above Fourier integral proceeds similarly to the derivation of the three-point spectral function. Upon introducing the light-cone momentum variables $p_i^{\pm} = \frac{- \omega_i \pm k_i}{2}$, the spectral function factorizes as
\begin{equation}
    \begin{split}
        \rho_{raaa} (p_1, p_2, p_3) = i \sin(3 \pi h_\phi) \sin(2 \pi h_\phi) \sin( \pi h_\phi) (\pi T)^{8 h_\phi} & \\
        \sum_{h, \bar{h}}  C^2 (h_\phi; h, \bar{h})  I_h (p_1^-, p_2^-, p_3^-) I_{\bar{h}} (p_1^+, p_2^+, p_3^+),&
   \end{split}
\end{equation}
where the fundamental integral $I_h$ is given by
\begin{equation} \label{basicintegral}
    \begin{split}
        I_h (p_1, p_2, p_3) =\int \left( \prod_{j=1}^3 d x_j \, e^{i p_j x_j} \right) \frac{z^{h} \, {}_2F_1 \left( h, h, 2h, z \right)}{\sinh^{2 h_\phi}(\pi T (x_1 - x_2)) \, \sinh^{2 h_\phi}(\pi T x_3)}.
    \end{split}
\end{equation}
To compute this, it is convenient to re-express the hypergeometric function via its Mellin-Barnes integral representation
\begin{equation}
    {}_2F_1(a,b,c,z) = \frac{\Gamma(c)}{\Gamma(a) \Gamma(b)} \int_{-i \infty}^{i \infty} \frac{\, ds}{2 \pi i} \frac{\Gamma(a+s) \Gamma(b+s) \Gamma(-s)}{\Gamma(c+s)} (-z)^s,
\end{equation}
where the integration is carried along a vertical curve in the complex $s-$plane. By exploiting the explicit expression ($\ref{cross_ratio_mink}$) for the cross ratios in Lorentzian signature, the product of hyperbolic functions we have to integrate over in $(\ref{basicintegral})$ becomes
\begin{equation}
    \frac{1}{\sinh^{2 h_\phi - h -s}(\pi T (x_1-x_2))} \frac{1}{\sinh^{2 h_\phi - h -s}(\pi T x_3)} \frac{1}{\sinh^{h + s}(\pi T (x_1-x_3))} \frac{1}{\sinh^{h + s}(\pi T x_2)}.
\end{equation}
We can disentangle the arguments of these hyperbolic functions by introducing the change of variables $u = x_1 - x_2$, $v = x_1 - x_3$ enforced via the identities $\int d u \, \delta(u - x_1 + x_2)$, $\int d v \, \delta(v - x_1 + x_3)$. Integration over $x_2$, $x_3$, $u$ and $v$ yields
\begin{equation} \label{inte_spect}
    \begin{split}
        I_h (p_1, p_2, p_3) = & \frac{2^{4 (h_\phi-1)}}{(\pi T)^4} \frac{\Gamma(2h)}{\Gamma^2(h)} \int_{-i \infty}^{i \infty}  \frac{\, ds}{2 \pi i} \, \frac{\Gamma(-s)}{\Gamma(2h+s) \Gamma^2(2 h_\phi - h -s)} \,  e^{i \pi s} \\
        & \int \frac{\, d \omega}{2 \pi} \, \Gamma \left( \frac{h +s}{2} + \frac{i (p_1 + p_2 + \omega)}{2 \pi T} \right) \Gamma \left( \frac{h +s}{2} - \frac{i (p_1 + p_2 + \omega)}{2 \pi T} \right) \\
        & \qquad \quad \text{                                             } \Gamma \left( \frac{h +s}{2} + \frac{i \omega}{2 \pi T} \right) \Gamma \left( \frac{h +s}{2} - \frac{i \omega}{2 \pi T} \right) \\
        & \qquad \quad \text{                                          } \Gamma \left( \frac{2h_\phi - h - s}{2} + \frac{i (p_1 +\omega)}{2 \pi T} \right) \Gamma \left( \frac{2h_\phi - h - s}{2} -\frac{i (p_1 +\omega)}{2 \pi T} \right) \\
        & \qquad \quad \text{                                          } \Gamma \left( \frac{2h_\phi - h - s}{2} + \frac{i (p_3 - \omega)}{2 \pi T} \right) \Gamma \left( \frac{2h_\phi - h - s}{2} - \frac{i (p_3 - \omega)}{2 \pi T} \right).
    \end{split}
\end{equation}
As outlined in the comment below equation $(\ref{basic_integral})$, this result can be analytically continued to (almost) any value in the $(2h_\phi - h - s)$- and $(h+s)-$planes. 

The integration variable $\omega$ originates from the Fourier transform of the delta function
\begin{equation}
    \delta(x) = \int \frac{\, d \omega}{2 \pi} e^{-i \, \omega \, x},
\end{equation}
and it represents the frequency of all possible intermediate states contributing to the four-point function. Although the corresponding integration path lies along the real axis, it can be trivially extended to the complex $\omega-$plane. We can then evaluate the $\omega-$integral by considering a contour closing in the lower half plane. In such a case, the poles coming from the Gamma functions are:
\begin{equation}
    \begin{split}
        1) \quad \omega & = - (p_1 + p_2) -2 \pi i T \left( n + \frac{h+s}{2} \right), \\
        2) \quad \omega & = - 2 \pi i T \left( \frac{h+s}{2} + n \right), \\ 
        3) \quad \omega & = - p_1 - 2 \pi i T \left( \frac{2h_\phi - h - s}{2} +n \right), \\
        4) \quad \omega & = p_3 - 2 \pi i T \left( \frac{2h_\phi - h - s}{2} +n \right).
    \end{split}
\end{equation}
Computing the corresponding residues and performing the $s-$integral in $(\ref{inte_spect})$ yields
\begin{equation} \label{eq:4pt_mom_space}
    \begin{split}
          & \rho_{raaa} (p_1, p_2, p_3) = i \sin(3 \pi h_\phi) \sin(2 \pi h_\phi) \sin( \pi h_\phi) (2 \pi T)^{8 (h_\phi -1)} \\
          &\sum_{h, \bar{h}} C^2 (h_\phi; h, \bar{h})
         \frac{\Gamma(2h)}{\Gamma^2(h)} \frac{\Gamma(2 \bar{h})}{\Gamma^2(\bar{h})} \sum_{n=0}^\infty \frac{(-1)^n}{n !} \sum_{i=1}^4 A_i \left( p_1^-, p_2^-, p_3^-; h,n \right) \, A_i \left( p_1^+, p_2^+, p_3^+; \bar{h},n \right),
        \end{split}
\end{equation}
where each kinematic factor $A_i(p_1, p_2, p_3;h,n)$ is related to a specific Meijer $G$-functions:
\begin{equation} \label{eq:A_1}
    \begin{split}
        & A_1 = \Gamma \left( -n + \frac{i \left(p_1 + p_2 \right) }{2 \pi T}  \right) \Gamma \left( h_\phi + n-\frac{i  p_2 }{2 \pi T} \right) \Gamma \left( h_\phi+n-\frac{i \left(p_1 + p_2 + p_3 \right)}{2 \pi T} \right)  \\ 
        & G_{4,4}^{3,2} 
        \begin{pmatrix}
            \frac{1}{2} - h - n, \quad \frac{1}{2} - h - n + \frac{i(p_1+p_2)}{2 \pi T}, \quad 2 h_\phi - h - \frac{1}{2}, \quad 2 h_\phi - h - \frac{1}{2}, \\
            & & e^{i \pi} \\
            -\frac{1}{2}, \quad h_\phi - h -\frac{1}{2} -n +\frac{i p_2}{2 \pi T}, \quad h_\phi -h - \frac{1}{2} - n + \frac{i \left(p_1 + p_2 + p_3 \right)}{2 \pi T}, \quad \frac{1}{2}-2h,
        \end{pmatrix},
    \end{split}
\end{equation}

\begin{equation}
    \begin{split}
        & A_2 = \Gamma \left( -n-\frac{i \left(p_1 + p_2 \right)}{2 \pi T} \right) \Gamma \left( h_\phi + n + \frac{i  p_1 }{2 \pi T} \right) \Gamma \left( h_\phi+n-\frac{i p_3}{2 \pi T} \right) \\ 
        & G_{4,4}^{3,2} 
        \begin{pmatrix}
            \frac{1}{2} - h - n, \quad \frac{1}{2} - h - n - \frac{i(p_1+p_2)}{2 \pi T}, \quad 2 h_\phi - h - \frac{1}{2}, \quad 2 h_\phi - h - \frac{1}{2}, \\
            & & e^{i \pi} \\
            -\frac{1}{2}, \quad h_\phi - h -\frac{1}{2} -n - \frac{i p_1}{2 \pi T}, \quad h_\phi -h - \frac{1}{2} - n + \frac{i p_3}{2 \pi T}, \quad \frac{1}{2}-2h,
        \end{pmatrix},
    \end{split}
\end{equation}

\begin{equation}
    \begin{split}
        & A_3 = \Gamma \left( h_\phi +n+\frac{i p_2}{2 \pi T} \right) \Gamma \left(h_\phi +n - \frac{ip_1}{2 \pi T} \right) \Gamma \left( -n + \frac{i (p_1+p_3)}{2 \pi T} \right)\\ 
        & G_{4,4}^{3,2} 
        \begin{pmatrix}
            \frac{2 h_\phi - 2h +1}{2} + n + \frac{ip_2}{2 \pi T}, \quad  \frac{2 h_\phi - 2h +1}{2} + n - \frac{ip_1}{2 \pi T}, \quad  \frac{4 h_\phi - 2h -1}{2}, \quad \frac{4 h_\phi - 2h -1}{2}, \\
            & & e^{i \pi} \\
            -\frac{1}{2}, \quad 2h_\phi - h -\frac{1}{2} +n, \quad 2h_\phi -h - \frac{1}{2} + n - \frac{i (p_1+p_3)}{2 \pi T}, \quad \frac{1}{2}-2h,
        \end{pmatrix},
    \end{split}
\end{equation}

\begin{equation} \label{eq:A_4}
    \begin{split}
        & A_4=\Gamma \left( h_\phi +n+\frac{i (p_1+p_2+p_3)}{2 \pi T} \right) \Gamma \left(h_\phi +n + \frac{ip_3}{2 \pi T} \right) \Gamma \left( -n - \frac{i (p_1+p_3)}{2 \pi T} \right) \\ 
        & G_{4,4}^{3,2} 
        \begin{pmatrix}
            \frac{2 h_\phi - 2h +1}{2} + n + \frac{i(p_1+p_2+p_3)}{2 \pi T}, \quad \frac{2 h_\phi - 2h +1}{2} + n + \frac{ip_3}{2 \pi T}, \quad \frac{4 h_\phi - 2h -1}{2}, \quad \frac{4 h_\phi - 2h -1}{2}, \\
            & & e^{i \pi} \\
            -\frac{1}{2}, \quad 2h_\phi - h -\frac{1}{2} +n + \frac{i(p_1+p_3)}{2 \pi T}, \quad 2h_\phi -h - \frac{1}{2} + n, \quad \frac{1}{2}-2h,
        \end{pmatrix}.
    \end{split}
\end{equation}
To verify the convergence of this spectral function at large $n$, we analyze the asymptotic scaling of its constituent factors. For $n \gg 1$, Gamma functions with arguments growing as $(n+a)$ scale via Stirling's approximation as $\Gamma(n+a) \sim n! \, n^{a-1}$, contributing a factorial growth factor. Conversely, Gamma functions whose arguments scale as $(-n+a)$ behave as $\Gamma(-n+a) \sim (-1)^n \pi / (\sin(\pi a) n!)$, providing an inverse factorial suppression. On the other hand, the upper and lower parameters of the Meijer $G$-functions shift linearly with $-n$; their large-$n$ asymptotics yield an additional suppression of order $\mathcal{O}(1/(n!)^2)$. Combining these competing effects, every individual term in the sum ($\ref{eq:4pt_mom_space}$) is suppressed by an overall inverse power of factorials at large $n$. The infinite series over $n$ is therefore absolutely convergent.

Although a further reduction of $\rho_{raaa}$ into a closed-form expression remains elusive, this formulation possesses an appealing structural layout: it cleanly mirrors the position-space OPE structure while making the infinite sequence of physical poles manifest. Indeed, we can map out the pole structure of the result by imposing the conditions $(\ref{eq:poles_cond})$, valid for any Meijer $G$-function. Specifically, for fixed values of $h$ and $n$, these constraints read
\begin{equation}
    \begin{split}
        1-h -n \notin \mathbb{N}, \qquad & h -2h_\phi -n \notin \mathbb{Z}, \qquad \pm \frac{i(p_1^-+p_3^-)}{2 \pi T} \notin\mathbb{Z}, \\
        1-h-n + \frac{i(p_1^-+p_2^-)}{2 \pi T} & \notin \mathbb{N}, \qquad -2 h_\phi +h -n\pm\frac{i(p_1^-+p_3^-)}{2 \pi T} \notin \mathbb{Z}, \\
        1 - h_\phi \pm \frac{i p_i^-}{2 \pi T} & \notin \mathbb{N}, \qquad h + n -h_\phi - \frac{ip_2^-}{2 \pi T}  \notin \mathbb{Z}, \qquad (i=1,2,3)\\
         h + n -h_\phi + \frac{ip_1^-}{2 \pi T} &\notin \mathbb{Z}, \qquad h + n -h_\phi - \frac{ip_3^-}{2 \pi T} \notin \mathbb{Z}, \\
        1 - h_\phi \pm \frac{i (p_1^-+p_2^-+p_3^-)}{2 \pi T} & \notin \mathbb{N}, \qquad h + n -h_\phi - \frac{i(p_1^-+p_2^-+p_3^-)}{2 \pi T} \notin \mathbb{Z},
    \end{split}
\end{equation}
along with symmetric expressions obtained via $h \to \bar{h}$, $p_i^- \to p_i^+$. 

As in the case of the three-point spectral function, a corresponding holographic computation of $\rho_{raaa}$ was carried out in~\cite{Loganayagam:2023}, where a different basis of four-point correlators was employed. Accordingly, we can compare the pole structure of the two results, which will differ precisely because of the choice of basis.

\section{Example: Ising model} \label{sec:example}

Thus far we have considered an abstract 2D CFT and we have analyzed the correlators of generic scalar primary operators. While this provides a general framework, it is also instructive to specialize the discussion to a concrete example. To this end, we examine the case in which the thermal bath is taken to be the 2D Ising model. This choice is motivated by the fact that it is one of the simplest nontrivial cases in which all the relevant data are explicitly known \cite{Mathieu:1997}.

The primary scalar operators appearing in the Ising model are the identity operator $I$, the spin operator $\sigma$ and the thermal operator $\epsilon$. The corresponding conformal dimensions are 
\begin{equation}\label{eq:data1}
    h_I = 0, \qquad h_\sigma = \frac{1}{16}, \qquad h_\epsilon = \frac{1}{2}.
\end{equation}
Moreover, the only non-zero structure constants involving primary operators are
\begin{equation}\label{eq:structure_const}
    C_{III}= C_{\epsilon \epsilon I} = C_{\sigma \sigma I} = 1, \qquad C_{\sigma \sigma \epsilon} = \frac{1}{2}.
\end{equation}
From the above CFT data it follows that the only non-trivial correlators involving only identical operators are $\langle \sigma_1 \sigma_2 \rangle$, $\langle \epsilon_1 \epsilon_2 \rangle$, $\langle \sigma_1 \sigma_2 \sigma_3 \sigma_4 \rangle$ and $\langle \epsilon_1 \epsilon_2 \epsilon_3 \epsilon_4 \rangle$. Since the OPE decomposition of the correlator $\langle \epsilon_1 \epsilon_2 \epsilon_3 \epsilon_4 \rangle$ involves only the identity block, it can be factorized into the product of two disconnected two-point functions. We neglect this degenerate case in the following.

We can immediately specialize the result ($\ref{integral_1}$) for the two-point spectral function to the Ising model by setting either $h_\phi \equiv  h_\sigma$ or $h_\phi \equiv h_\epsilon$:
\begin{equation}
    \begin{split}
        \rho_{ra}^{\sigma}(p) & = \frac{- i \, \sqrt{2 - \sqrt{2}}}{(2 \pi T)^{\frac{7}{4}} \, \Gamma^2 \left( \frac{1}{8} \right)} \, \prod_{\delta = \pm} \Gamma \left( \frac{1}{16} + \frac{i p^\delta}{2 \pi T} \right) \, \Gamma \left( \frac{1}{16} - \frac{i p^\delta}{2 \pi T} \right), \\
        \rho_{ra}^\epsilon(p) & = 0.
    \end{split}
\end{equation}
The two-point spectral function associated with $\langle \epsilon_1 \epsilon_2 \rangle$ vanishes as $h_\epsilon$ is half-integer.

Similarly, the case of the four-point spectral function does not simplify significantly compared to the general situation. Although an exact expression for the four-point function is known
\begin{equation}
    \langle \sigma(z_1) \sigma(z_2) \sigma (z_3) \sigma (z_4) \rangle = \frac{\sqrt{z \bar{z}} + \sqrt{(1-z)(1- \bar{z})}}{\left( z \bar{z} (1-z) (1 - \bar{z}) \right)^{\frac{1}{8}}},
\end{equation}
it is not possible to find an equally explicit result for the corresponding finite-temperature correlator in momentum space. Hence, the most convenient expression for the four-point spectral function $\rho_{raaa}^\sigma$ relies on the general formulas $(\ref{eq:4pt_mom_space})$-$(\ref{eq:A_4})$. By exploiting the CFT data ($\ref{eq:data1}$) and ($\ref{eq:structure_const}$) we can explicitly write
\begin{equation}
    \begin{split}
        & \rho_{raaa}^\sigma (p_1, p_2, p_3) = \frac{ i \left( \sqrt{2}-\sqrt{4-2\sqrt{2}} \,\right)}{32 \, \pi^2 \, (2 \pi T)^{\frac{15}{2}}} \sum_{n=0}^\infty \frac{(-1)^n}{n !} \sum_{i=1}^4 \prod_{\delta=\pm} A_i \left( p_1^\delta, p_2^\delta, p_3^\delta; \frac{1}{2},n \right) \\
        & - \frac{3-2\sqrt{2}}{ 2 \, (2 \pi T)^{\frac{7}{2}} \, \Gamma^4 \left( \frac{1}{8} \right)} \prod_{\delta = \pm} (2 \pi) \, \delta \left( p_1^\delta + p_2^\delta \right) \, \prod_{j=1,3} \Gamma \left( \frac{1}{16} + \frac{i p_j^\delta}{2 \pi T} \right) \, \Gamma \left( \frac{1}{16} - \frac{i p_j^\delta}{2 \pi T} \right).
    \end{split}
\end{equation}
The second term, associated with the identity block, corresponds simply to the Fourier transform of two disconnected two-point functions $\langle \sigma_1 \sigma_2\rangle \langle \sigma_3 \sigma_4 \rangle$. In contrast, the first term arises from the global conformal block associated with the thermal operator $\epsilon$ and represents a genuinely non-trivial contribution to the four-point spectral function. As it retains the general structure previously derived, the coefficients $A_i \left( p_1^\delta, p_2^\delta, p_3^\delta; \frac{1}{2}, n \right)$ can be determined from the general expressions $(\ref{eq:A_1})$-$(\ref{eq:A_4})$ by setting $h_\phi \equiv h_\sigma$.

\section{Conclusion} \label{sec_conclusion}

Initially motivated by the study of the effective dynamics of an open CFT, in this paper we presented the momentum-space representation of the Lorentzian two-, three- and four-point causal response functions of an arbitrary 2D CFT. While the two- and three-point correlators can be straightforwardly analytically continued from Euclidean to Lorentzian signature, the same does not hold for the four-point function. In principle, a full real-time reconstruction requires tracking 24 different time orderings. While a complete recovery of all orderings remains out of reach within this framework, we have demonstrated that it is entirely possible to isolate the specific time-ordered commutators necessary to construct the physical response function.

There are several possible generalizations of our work. We have focused our analysis on the case of 2D CFTs, as they provide the simplest non-trivial laboratory. However, there is no conceptual barrier preventing a generalization of our results to higher dimensions. Furthermore, our approach could be naturally extended to thermal correlators involving non-identical scalar fields, as well as fields with non-zero spin.

In principle, we could apply the same logic presented in this paper to determine higher-order contributions to the effective dynamics of an open quantum system. This would require computing higher-order $n-$point correlators, which would depend on $\frac{n(n-3)}{2}$ independent cross-ratios. However, as $n$ increases, the computations would become increasingly complex, leaving little hope of obtaining closed-form results.

Finally, in this paper we focused on the investigation of open quantum systems described by CFTs. Although they are immensely valuable for understanding critical phenomena and phase transitions, we would ultimately like to analyze more general scenarios involving arbitrary QFTs. However, in these cases the perturbative approach leads to effective dynamics plagued by non-locality. Addressing this issue would require moving beyond perturbation theory and developing new methods to study non-perturbative effective theories in QFT.

\section*{Acknowledgments}\addcontentsline{toc}{section}{Acknowledgments}

I would like to thank my Ph.D advisor, Prof. Mukund Rangamani, for being the first to introduce me to the problem addressed in this paper, for guiding me through its preparation and for providing many valuable suggestions.

\newpage
\bibliographystyle{JHEP}
\bibliography{opencfts}

@article{Dolan:2000,
  author        = {Dolan, F.A. and Osborn, H.},
  title         = {{Conformal Four Point Functions and Operator Porduct Expansion}},
  eprint        = {hep-th/0011040},
  archiveprefix = {arXiv},
  doi           = {10.1016/S0550-3213(01)00013-X},
  journal       = {Nucl.Phys. B},
  volume        = {599},
  pages         = {459-496},
  year          = {2000}
}

@article{Hogervorst:2013,
  author        = {Hogervorst, M. and Rychkov S.},
  title         = {{Radial coordinates for conformal blocks}},
  eprint        = {1303.1111 [hep-th]},
  archiveprefix = {arXiv},
  doi           = {},
  journal       = {Physical Review D},
  volume        = {87},
  pages         = {},
  year          = {2013}
}

@article{Becker:2014,
  author        = {Becker, M. and Cabrera, Y. and Su, N.},
  title         = {{Finite-temperature three-point function in 2D CFT}},
  eprint        = {1407.3415 [hep-th]},
  archiveprefix = {arXiv},
  doi           = {10.1007/JHEP09(2014)157},
  journal       = {JHEP},
  volume        = {09},
  pages         = {157},
  year          = {2014}
}

@article{Haehl:2019,
  author        = {Haehl, F. M. and Loganayagam, R. and Narayan, P. and M. Rangamani},
  title         = {{Classification of out-of-time-order correlators}},
  eprint        = {1701.02820 [hep-th]},
  archiveprefix = {arXiv},
  doi           = {10.21468/SciPostPhys.6.1.001},
  journal       = {SciPost Phys.},
  volume        = {6},
  pages         = {1},
  year          = {2019}
}

@article{Hartman:2019,
  author        = {Hartman, T. and Jain, S. and Kundu, S.},
  title         = {{Causality Constraints in Conformal Field Theory}},
  eprint        = {1509.00014 [hep-th]},
  archiveprefix = {arXiv},
  doi           = {10.1007/JHEP05(2016)099},
  journal       = {JHEP},
  volume        = {5},
  pages         = {99},
  year          = {2016}
}

@article{Chaudhuri:2019,
  author        = {Chaudhuri, S. and Chowdhury, C. and Loganayagam, R.},
  title         = {{Spectral Representation of Thermal OTO Correlators}},
  eprint        = {1810.03118 [hep-th]},
  archiveprefix = {arXiv},
  doi           = {10.1007/JHEP02(2019)018},
  journal       = {JHEP},
  volume        = {2},
  pages         = {18},
  year          = {2019}
}

@article{Perlmutter:2015,
  author        = {Perlmutter, E.},
  title         = {{Virasoro conformal blocks in closed form}},
  eprint        = {1502.07742 [hep-th]},
  archiveprefix = {arXiv},
  doi           = {10.1007/JHEP08(2015)088},
  journal       = {JHEP},
  volume        = {8},
  pages         = {15},
  year          = {2015}
}

@article{Pappadopulo:2012,
  author        = {Pappadopulo, D. and Rychkov, S. and Espin, J. and Rattazzi, R.},
  title         = {{OPE Convergence in Conformal Field Theory}},
  eprint        = {1208.6449 [hep-th]},
  archiveprefix = {arXiv},
  doi           = {10.1103/PhysRevD.86.105043},
  journal       = {Phys.Rev.D},
  volume        = {86},
  pages         = {},
  year          = {2012}
}

@article{Qiao:2022,
  author        = {Qiao, J.},
  title         = {{On the Wick rotation of the Four-point functions in Conformal Field Theory}},
  eprint        = {2209.00285 [hep-th]},
  archiveprefix = {arXiv},
  doi           = {},
  journal       = {},
  volume        = {},
  pages         = {},
  year          = {2022}
}

@article{Horowitz:2000,
  author        = {Horowitz, G. T. and Hubeny,V. E.},
  title         = {{Quasinormal modes of AdS black holes and the approach to thermal equilibrium}},
  eprint        = {hep-th/9909056},
  archiveprefix = {arXiv},
  doi           = {10.1103/PhysRevD.62.024027},
  journal       = {Phys.Rev.D},
  volume        = {62},
  pages         = {},
  year          = {2000}
}

@article{Haehl:2017_2,
  author        = {Haehl, F. M. and Loganayagam, R. and Rangamani, M.},
  title         = {{Schwinger-Keldysh formalism. Part II: thermal equivariant cohomology}},
  eprint        = {1610.01941 [hep-th]},
  archiveprefix = {arXiv},
  doi           = {10.1007/JHEP06(2017)070},
  journal       = {JHEP},
  volume        = {06},
  pages         = {070},
  year          = {2017}
}

@article{Maldacena:2016,
  author        = {Maldacena, J. and Shenker, S. H. and Stanford, D.},
  title         = {{A bound on chaos}},
  eprint        = {1503.01409 [hep-th]},
  archiveprefix = {arXiv},
  doi           = {10.1007/JHEP08(2016)106},
  journal       = {JHEP},
  volume        = {08},
  pages         = {106},
  year          = {2016}
}

@article{Bautista:2020,
  author        = {Bautista, T. and Godazgar, H.},
  title         = {{Lorentzian CFT 3-point functions in momentum space}},
  eprint        = {1908.04733 [hep-th]},
  archiveprefix = {arXiv},
  doi           = {10.1007/JHEP01(2020)142},
  journal       = {JHEP},
  volume        = {01},
  pages         = {142},
  year          = {2020}
}

@article{Gubser:1997,
  author        = {Gubser, S.S.},
  title         = {{Absorption of photons and fermions by black holes in four dimensions}},
  eprint        = {hep-th/9706100},
  archiveprefix = {arXiv},
  doi           = {10.1103/PhysRevD.56.7854},
  journal       = {Phys.Rev.D},
  volume        = {56},
  pages         = {7854-7868},
  year          = {1997}
}

@article{Roberts:2015,
  author        = {Roberts, D. A. and Stanford, D.},
  title         = {{Two-dimensional conformal field theory and the butterfly effect}},
  eprint        = {1412.5123 [hep-th]},
  archiveprefix = {arXiv},
  doi           = {10.1103/PhysRevLett.115.131603},
  journal       = {Phys.Rev.Lett.},
  volume        = {115},
  pages         = {},
  year          = {2015}
}

@article{Feynman:1963,
  author        = {Feynman, R. and Vernon, F.},
  title         = {{The theory of a general quantum system interacting with a linear dissipative system}},
  eprint        = {},
  archiveprefix = {},
  doi           = {10.1016/0003-4916(63)90068-X},
  journal       = {Annals Phys.},
  volume        = {24},
  pages         = {118-173},
  year          = {1963}
}

@article{Haehl:2017,
  author        = {Haehl, F. M. and Loganayagam, R. and Rangamani, M.},
  title         = {{Schwinger-Keldysh formalism I: BRST symmetries and superspace}},
  eprint        = {1610.01940 [hep-th]},
  archiveprefix = {arXiv},
  doi           = {10.1007/JHEP06(2017)069},
  journal       = {JHEP},
  volume        = {06},
  pages         = {069},
  year          = {2017}
}

@article{Pelliconi:2024,
  author        = {Pelliconi, P. and Sonner, J.},
  title         = {{The Influence Functional in open holography:
entanglement and Renyi entropies}},
  eprint        = {2310.13047 [hep-th]},
  archiveprefix = {arXiv},
  doi           = {10.1007/JHEP06(2024)185},
  journal       = {JHEP},
  volume        = {06},
  pages         = {185},
  year          = {2024}
}

@article{Birmingham:2002,
  author        = {Birmingham, Danny and Sachs, I. and Solodukhin, S.},
  title         = {{Conformal field theory interpretation of black hole quasinormal modes}},
  eprint        = {hep-th/0112055},
  archiveprefix = {arXiv},
  doi           = {10.1103/PhysRevLett.88.151301},
  journal       = {Phys.Rev.Lett.},
  volume        = {88},
  pages         = {},
  year          = {2002}
}

@article{Son:2002,
  author        = {Son, D.T. and Starinets, A.O.},
  title         = {{Minkowski space correlators in AdS / CFT correspondence: Recipe and applications}},
  eprint        = {hep-th/0205051},
  archiveprefix = {arXiv},
  doi           = {10.1088/1126-6708/2002/09/042},
  journal       = {JHEP },
  volume        = {09},
  pages         = {042},
  year          = {2002}
}

@article{Loganayagam:2023,
  author        = {Loganayagam, R. and Rangamani, M. and Virrueta, J.},
  title         = {{Holographic open quantum systems: toy models and analytic properties of thermal correlators}},
  eprint        = {2211.07683 [hep-th]},
  archiveprefix = {arXiv},
  doi           = {10.1007/JHEP03(2023)153},
  journal       = {JHEP},
  volume        = {03},
  pages         = {153},
  year          = {2023}
}

@article{Chandan:2020,
  author        = {Chandan, J. and Loganayagam, R. and Rangamani, M.},
  title         = {{Open quantum systems and Schwinger-Keldysh holograms}},
  eprint        = {2004.02888 [hep-th]},
  archiveprefix = {arXiv},
  doi           = {10.1007/JHEP07(2020)242},
  journal       = {JHEP},
  volume        = {07},
  pages         = {242},
  year          = {2020}
}

@article{Haehl:2017_kms,
  author        = {Haehl, F. and Loganayagam, R. and Narayan, P. and Nizami, A. A. and Rangamani, M.},
  title         = {{Thermal out-of-time-order correlators, KMS relations, and spectral functions}},
  eprint        = {1706.08956 [hep-th]},
  archiveprefix = {arXiv},
  doi           = {10.1007/JHEP12(2017)154},
  journal       = {JHEP},
  volume        = {12},
  pages         = {154},
  year          = {2017}
}

@article{Manenti:2020,
  author        = {Manenti, A.},
  title         = {{Thermal CFTs in momentum space}},
  eprint        = {1905.01355 [hep-th]},
  archiveprefix = {arXiv},
  doi           = {10.1007/JHEP01(2020)009},
  journal       = {JHEP},
  volume        = {01},
  pages         = {009},
  year          = {2020}
}

@article{Poland:2019,
  author        = {Poland, David and Rychkov, Slava and Vichi, Alessandro},
  title         = {{The Conformal Bootstrap: Theory, Numerical Techniques, and Applications}},
  eprint        = {1805.04405 [hep-th]},
  archiveprefix = {arXiv},
  doi           = {10.1103/RevModPhys.91.015002},
  journal       = {Rev.Mod.Phys.},
  volume        = {91},
  pages         = {015002},
  year          = {2019}
}

@book{Haang:1992,
    author = {Haang, R.},
    title = {{Local Quantum Physics: Fields, Particles, Algebras}},
    publisher = {Springer},
    year = {1992}
}

@book{Breuer:2002,
    author = {Breuer, H. and Petruccione, F.},
    title = {{The theory of open quantum systems}},
    publisher = {Oxford University Press},
    year = {2002}
}

@article{Gfunction,
    author = {NIST},
    title         = {{NIST Digital Library of Mathematical Functions}},
    eprint        = {},
    archiveprefix = {},
    doi           = {},
    journal       = {https://dlmf.nist.gov/},
    volume        = {},
    pages         = {},
    year          = {Release 1.2.3 of 2025-03-15}
}

@article{Schwinger:1961,
    author = {Schwinger, Julian S.},
    title         = {{Brownian motion of a quantum oscillator}},
    eprint        = {},
    archiveprefix = {},
    doi           = {10.1063/1.1703727},
    journal       = {J.Math.Phys.},
    volume        = {2},
    pages         = {407-432},
    year          = {1961}
}

@article{Keldysh:1965,
    author = {Keldysh, L.V.},
    title         = {{Diagram Technique for Nonequilibrium Processes}},
    eprint        = {},
    archiveprefix = {},
    doi           = {},
    journal       = {Sov.Phys.JETP},
    volume        = {20},
    pages         = {1018-1026},
    year          = {1965}
}

@article{Chou:1985,
    author = {Chou, Kuang-chao and Su, Zhao-bin and Hao, Bai-lin and Yu, Lu},
    title         = {{Equilibrium and Nonequilibrium Formalisms Made Unified
}},
    eprint        = {},
    archiveprefix = {},
    doi           = {10.1016/0370-1573(85)90136-X},
    journal       = {Phys.Rept.},
    volume        = {118},
    pages         = {1-131},
    year          = {1985}
}

@article{Landsman:1987,
    author = {Landsman, N.P. and van Weert, C.G.},
    title         = {{Real and Imaginary Time Field Theory at Finite Temperature and Density}},
    eprint        = {},
    archiveprefix = {},
    doi           = {10.1016/0370-1573(87)90121-9},
    journal       = {Phys.Rept.},
    volume        = {145},
    pages         = {141},
    year          = {1987}
}

@article{Kamenev:2009,
    author = {Kamenev, Alex and Levchenko, Alex},
    title         = {{Keldysh technique and nonlinear sigma-model: Basic principles and applications}},
    eprint        = {0901.3586 [cond-mat.other]},
    archiveprefix = {arXiv},
    doi           = {10.1080/00018730902850504},
    journal       = {Adv.Phys.},
    volume        = {58},
    pages         = {197},
    year          = {2009}
}

@article{Caldeira:1983,
    author = {Caldeira, A.O. and Leggett, A.J.},
    title         = {{Path integral approach to quantum Brownian motion}},
    eprint        = {},
    archiveprefix = {},
    doi           = {10.1016/0378-4371(83)90013-4},
    journal       = {Physica A},
    volume        = {121},
    pages         = {587-616},
    year          = {1983}
}

@article{Grozdanov:2015,
    author = {Grozdanov, Sašo and Polonyi, Janos},
    title         = {{Viscosity and dissipative hydrodynamics from effective field theory}},
    eprint        = {1305.3670 [hep-th]},
    archiveprefix = {arXiv},
    doi           = {10.1103/PhysRevD.91.105031},
    journal       = {Phys.Rev.D},
    volume        = {91},
    pages         = {10},
    year          = {2015}
}

@article{Kubo:1957,
    author = {Kubo, Ryogo},
    title         = {{Statistical mechanical theory of irreversible processes. 1. General theory and simple applications in magnetic and conduction problems}},
    eprint        = {},
    archiveprefix = {},
    doi           = {10.1143/JPSJ.12.570},
    journal       = {J.Phys.Soc.Jap.},
    volume        = {12},
    pages         = {570-586},
    year          = {1957}
}

@article{Martin:1959,
    author = {Martin, Paul C. and Schwinger, Julian},
    title         = {{Theory of Many-Particle Systems. I}},
    eprint        = {},
    archiveprefix = {},
    doi           = {10.1143/JPSJ.12.570},
    journal       = {Phys. Rev.},
    volume        = {115},
    pages         = {1342–1373},
    year          = {1959}
}

@article{Gransee:2017,
    author = {Gransee, Michael and Pinamonti, Nicola and Verch, Rainer},
    title         = {{KMS-like Properties of Local Equilibrium States in Quantum Field Theory}},
    eprint        = {1508.05585 [math-ph]},
    archiveprefix = {arXiv},
    doi           = {10.1016/j.geomphys.2017.02.014},
    journal       = {J.Geom.Phys.},
    volume        = {117},
    pages         = {215-236},
    year          = {2017}
}

@article{Kravchuk:2021,
    author = {Kravchuk, Petr and Qiao, Jiaxin and Rychkov, Slava},
    title         = {{Distributions in CFT. Part II. Minkowski space}},
    eprint        = {2104.02090 [hep-th]},
    archiveprefix = {arXiv},
    doi           = {10.1007/JHEP08(2021)094},
    journal       = {JHEP},
    volume        = {08},
    pages         = {094},
    year          = {2021}
}

@article{Kravchuk:2020,
    author = {Kravchuk, Petr and Qiao, Jiaxin and Rychkov, Slava},
    title         = {{Distributions in CFT. Part I. Cross-ratio space}},
    eprint        = {2001.08778 [hep-th]},
    archiveprefix = {arXiv},
    doi           = {10.1007/JHEP05(2020)137},
    journal       = {JHEP},
    volume        = {05},
    pages         = {137},
    year          = {2020}
}

@article{Qiao:202222,
    author = {Qiao, Jiaxin},
    title         = {{Classification of Convergent OPE Channels for Lorentzian CFT Four-Point Functions}},
    eprint        = {2005.09105 [hep-th]},
    archiveprefix = {arXiv},
    doi           = {10.21468/SciPostPhys.13.4.093},
    journal       = {SciPost Phys.},
    volume        = {13},
    pages         = {4},
    year          = {2022}
}

@article{Haag:1967,
    author = {Haag, R. and Hugenholtz, N.M. and Winnink, M.},
    title         = {{On the Equilibrium states in quantum statistical mechanics}},
    eprint        = {},
    archiveprefix = {},
    doi           = {10.1007/BF01646342},
    journal       = {Commun.Math.Phys.},
    volume        = {5},
    pages         = {215-236},
    year          = {1967}
}

@article{Fitzpatrick:2017,
    author = {Fitzpatrick, A. Liam and Kaplan, Jared},
    title         = {{On the Late-Time Behavior of Virasoro Blocks and a Classification of Semiclassical Saddles}},
    eprint        = {1609.07153 [hep-th]},
    archiveprefix = {arXiv},
    doi           = {10.1007/JHEP04(2017)072},
    journal       = {JHEP},
    volume        = {04},
    pages         = {072},
    year          = {2017}
}

@article{Chang:2019,
    author = {Chang, Chi-Ming and Ramirez, David M. and Rangamani, Mukund},
    title         = {{Spinning constraints on chaotic large c CFTs}},
    eprint        = {1812.05585 [hep-th]},
    archiveprefix = {arXiv},
    doi           = {10.1007/JHEP03(2019)068},
    journal       = {JHEP},
    volume        = {03},
    pages         = {068},
    year          = {2019}
}

@article{Sieberer:2016,
    author = {Sieberer, L.M. and Diehl, S.},
    title         = {{Keldysh Field Theory for Driven Open Quantum Systems}},
    eprint        = {1512.00637 [cond-mat.quant-gas]},
    archiveprefix = {arXiv},
    doi           = {10.1088/0034-4885/79/9/096001},
    journal       = {Rept.Prog.Phys.},
    volume        = {79},
    pages         = {9},
    year          = {2016}
}

@article{Reyes:2026,
    author = {Reyes-Osorio, Felipe and Garcia-Gaitan, Federico and Strachan, David J. and Plechac, Petr and Clark, Stephen R.},
    title         = {{Schwinger–Keldysh nonperturbative field theory of open quantum systems beyond the Markovian regime: application to spin-boson and spin-chain-boson models}},
    eprint        = {2405.00765 [quant-ph]},
    archiveprefix = {arXiv},
    doi           = {10.1088/1361-6633/ae2888},
    journal       = {Rept.Prog.Phys.},
    volume        = {89},
    pages         = {1},
    year          = {2026}
}

@article{Maldacena:2017,
    author = {Maldacena, Juan and Simmons-Duffin, David and Zhiboedov, Alexander},
    title         = {{Looking for a bulk point}},
    eprint        = {1509.03612 [hep-th]},
    archiveprefix = {arXiv},
    doi           = {10.1007/JHEP01(2017)013},
    journal       = {JHEP},
    volume        = {01},
    pages         = {013},
    year          = {2017}
}

@article{Gillioz:2025,
    author = {Gillioz, Marc},
    title         = {{The momentum-space conformal bootstrap in 2d}},
    eprint        = {2502.21227 [hep-th]},
    archiveprefix = {arXiv},
    doi           = {},
    journal       = {},
    volume        = {},
    pages         = {},
    year          = {2025}
}

@article{Gillioz:2021,
    author = {Gillioz, Marc},
    title         = {{Conformal partial waves in momentum space}},
    eprint        = {2012.09825 [hep-th]},
    archiveprefix = {arXiv},
    doi           = {10.21468/SciPostPhys.10.4.081},
    journal       = {SciPost Phys.},
    volume        = {10},
    pages         = {4},
    year          = {2021}
}

@article{Gillioz:2020,
    author = {Gillioz, Marc and Lu, Xiaochuan and Luty, Markus A. and Mikaberidze, Guram},
    title         = {{Convergent Momentum-Space OPE and Bootstrap Equations in Conformal Field Theory}},
    eprint        = {1912.05550 [hep-th]},
    archiveprefix = {arXiv},
    doi           = {10.1007/JHEP03(2020)102},
    journal       = {JHEP},
    volume        = {03},
    pages         = {102},
    year          = {2020}
}

@article{Gillioz:20202,
    author = {Gillioz, Marc and Lu, Xiaochuan and Luty, Markus A. and Mikaberidze, Guram},
    title         = {{Conformal 3-point functions and the Lorentzian OPE in momentum space}},
    eprint        = {1909.00878 [hep-th]},
    archiveprefix = {arXiv},
    doi           = {10.1007/s00220-020-03836-8},
    journal       = {Commun.Math.Phys.},
    volume        = {379},
    pages         = {227-259},
    year          = {2020}
}

@article{Zurek:2003,
    author = {Zurek, Wojciech Hubert},
    title         = {{Decoherence, einselection, and the quantum origins of the classical}},
    eprint        = {quant-ph/0105127},
    archiveprefix = {arXiv},
    doi           = {10.1103/RevModPhys.75.715},
    journal       = {Rev.Mod.Phys.},
    volume        = {75},
    pages         = {715-775},
    year          = {2003}
}

@article{Agon:2018,
    author = {Agón, Cesar and Lawrence, Albion},
    title         = {{Divergences in open quantum systems}},
    eprint        = {1709.10095 [hep-th]},
    archiveprefix = {arXiv},
    doi           = {10.1007/JHEP04(2018)008},
    journal       = {JHEP},
    volume        = {04},
    pages         = {008},
    year          = {2018}
}

@article{Boyanovsky:2015,
    author = {Boyanovsky, D.},
    title         = {{Effective Field Theory out of Equilibrium: Brownian quantum fields}},
    eprint        = {1503.00156 [hep-ph]},
    archiveprefix = {arXiv},
    doi           = {10.1088/1367-2630/17/6/063017},
    journal       = {New J.Phys.},
    volume        = {17},
    pages         = {6},
    year          = {2015}
}

@article{Boyanovsky:2018,
    author = {Boyanovsky, D.},
    title         = {{Information loss in effective field theory: entanglement and thermal entropies}},
    eprint        = {1801.06840 [hep-th]},
    archiveprefix = {arXiv},
    doi           = {10.1103/PhysRevD.97.065008},
    journal       = {Phys.Rev.D},
    volume        = {97},
    pages         = {6},
    year          = {2018}
}

@article{Baidya:2017,
    author = {Baidya, Avinash and Jana, Chandan and Loganayagam, R. and Rudra, Arnab},
    title         = {{Renormalization in open quantum field theory. Part I. Scalar field theory}},
    eprint        = {1704.08335 [hep-th]},
    archiveprefix = {arXiv},
    doi           = {10.1007/JHEP11(2017)204},
    journal       = {JHEP},
    volume        = {11},
    pages         = {204},
    year          = {2017}
}

@book{Kamenev_2011,
    author = {Kamenev, Alex},
    title  = {{Field Theory of Non-Equilibrium Systems}},
    publisher = {Cambridge University Press},
    year = {2011}
}

@article{Zamolodchikov:1987,
    author = {Zamolodchikov, Al. B.},
    title         = {{Conformal symmetry in two-dimensional space: Recursion representation of conformal block}},
    eprint        = {},
    archiveprefix = {},
    doi           = {10.1007/BF01022967},
    journal       = {Theor.Math.Phys.},
    volume        = {73},
    pages         = {1},
    year          = {1987}
}

@book{Polchinski:2007,
    author        = {Polchinski, J.},
    title         = {String theory. Vol. 1: An introduction to the bosonic string},
    publisher     = {Cambridge
Monographs on Mathematical Physics, Cambridge University Press},
    year          = {2007}
}

@book{Mathieu:1997,
    author        = {Mathieu, Pierre and Sénéchal, David and Di Francesco, Philippe},
    title         = {Conformal Field Theory},
    publisher     = {Springer},
    year          = {1997}
}

@article{Iliesiu:2018,
    author = {Iliesiu, Luca and Koloğlu, Murat and Mahajan, Raghu and Perlmutter, Eric and Simmons-Duffin, David},
    title         = {{The Conformal Bootstrap at Finite Temperature}},
    eprint        = {1802.10266 [hep-th]},
    archiveprefix = {arXiv},
    doi           = {10.1007/JHEP10(2018)070},
    journal       = {JHEp},
    volume        = {10},
    pages         = {070},
    year          = {2018}
}

@article{Costa:2012,
    author = {Costa, Miguel S. and Gonçalves, Vasco David Fonseca and Penedones, Joao},
    title         = {{Conformal Regge theory}},
    eprint        = {1209.4355 [hep-th]},
    archiveprefix = {arXiv},
    doi           = {10.1007/JHEP12(2012)091},
    journal       = {JHEp},
    volume        = {12},
    pages         = {091},
    year          = {2012}
}

\end{document}